\documentclass[pdftex,11pt,onecolumn,epjc3]{svjour3}
\RequirePackage[T1]{fontenc}

\smartqed  % flush right qed marks, e.g. at end of proof
\usepackage[english]{babel}
\usepackage[latin1]{inputenc}
\usepackage{amsmath}
\usepackage{amsbsy}

\usepackage[colorlinks=true,linkcolor=blue,citecolor=blue]{hyperref}
\usepackage{appendix}
\usepackage{xcolor}
\usepackage{amsfonts}
\usepackage{amssymb}
\usepackage{graphicx}
\usepackage{amsmath}
\usepackage{float}
\usepackage[english]{babel}
\usepackage[latin1]{inputenc}
\usepackage{ulem}
\usepackage{amsmath}

\usepackage[colorlinks=true,linkcolor=blue,citecolor=blue]{hyperref}
\usepackage{appendix}
\usepackage{xcolor}

\usepackage[colorlinks=true,linkcolor=blue,citecolor=blue]{hyperref}
\usepackage{appendix}
\usepackage{xcolor}

\journalname{Eur. Phys. J. C}

\begin{document}
\title{Remarks on propagating waves in non-linear vacuum electrodynamics}

\author{M. A. P\'erez-Garc\'ia\thanksref{e1,addr1}
        \and
A. P\'erez Mart\'inez\thanksref{e2,addr1}    \and
E. Rodr\'iguez Querts\thanksref{e3,addr2} %etc.
}

%\thankstext[$\star$]{t1}{Thanks to the title}
\thankstext{e1}{e-mail: mperezga@usal.es}
\thankstext{e2}{e-mail:aurorapm1961@usal.es}
\thankstext{e3}{e-mail:elizabeth@icimaf.cu}
\institute{Department of Fundamental Physics, University of Salamanca, Plaza de la Merced s/n 37008, Spain\label{addr1} \and
\noindent Departamento de F\'isica Te\'orica, Instituto de Cib\'ernetica Matem\'atica y F\'isica (ICIMAF),
Calle E esq 15 No. 309 Vedado, La Habana, 10400, Cuba  \label{addr2}
        %  \and
     %     \emph{Present Address:} Street, City, Country\label{addr3}
}
\date{\today}
%Received: date / Accepted: date}
\maketitle

\begin{abstract}
Using the quadratic expansion in the photon fields of Euler-Heisenberg (EH) non-linear electrodynamics (NLED) Lagrangian model we study relevant vacuum properties in a scenario involving the propagation of a photon probe in the presence of a background constant and static magnetic field, ${\bf B_e}$. We compute the gauge invariant, symmetric and conserved energy-momentum tensor (EMT) and  angular momentum tensor (AMT) for arbitrary magnetic field strength  using the Hilbert method under the soft-photon approximation.  We discuss how the presence of  magneto-electric terms in the EH Lagrangian is a source of anisotropy,  induce the non-zero trace in the EMT  and leads to  differences between EMT calculated by the  Hilbert or Noether method.
From the Hilbert EMT we analyze some quantities of interest such as the energy density, pressures, Poynting vector, and angular momentum vector, comparing and discussing the differences with respect to  the improved Noether method. The magnetized vacuum properties are also studied showing that a photon effective magnetic moment can be defined for different polarization modes. The calculations are done in terms of derivatives of the two scalar invariants of electrodynamics, hence, extension to other NLED Lagrangian is straightforward.
We discuss further physical implications and experimental strategies to test magnetization, photon pressure, and effective magnetic moment.

\end{abstract}

\section{Introduction}

Vacuum in NLED theories is far from being trivial. For example, in Quantum Electrodynamics (QED)
the vacuum polarization effects lead to effective interaction terms that are non-linear in the electric and magnetic fields, generating, among other phenomena, light-by-light scattering, first calculated in the low energy limit by Euler and
Kockel \cite{HEKockel} in 1935 and later completed by Karplus and Neuman \cite{lbyl} in 1951. As recently claimed photon-photon scattering may be
 testable with modern accelerators, see a recent compilation  from ATLAS results in \cite{atlas}\cite{Baier:2018vso}.
Switching on the interaction leads to excitations  creating virtual electron-positron pairs. %by exciting an electron from the lower to the upper continuum with subsequent deexcitation. These perturbations affect all quantities calculated in QED. In this way this
This theory conceives that fluctuations can give rise to very interesting properties allowing to describe vacuum as a magnetized medium \cite{Adler}. One of the quoted consequences is the possible existence of  {\it birefringence}, by which electromagnetic (EM) waves propagating parallel or perpendicular to a constant electric or magnetic background field in fixed direction display different propagation speeds. 
Great experimental effort has been undertaken in the task of detection, like that by Paulus and coworkers using an x-ray free-electron laser in EuXFEL \cite{jena}, but to present date this phenomenon has not been experimentally found yet, as it involves tiny effects \cite{melrose} due to the non-linear coupling $\xi=\frac{8\alpha^2 \hbar^3}{45m_e^4c^5}\sim \frac{8\alpha}{45B_c^2}\sim 6.7\times 10^{-30}$ $\rm m^3/J$ where $B_c=\frac{m_e^2 c^2}{e\hbar}=4.4\times 10^{13}$ G  is the critical magnetic field.

Alternatively, this non-detection has been interpreted as a possible  manifestation  of new physics. One of the most relevant experimental setups to capture the effect of birefringence is the Polarization of Vacuum with Laser (PVLAS)\cite{Ejlli:2020yhk} that proposed  axions \cite{Cadene} being produced from photon decay and thus novel effects being responsible for the null experimental signal, contrary to expectation. To date, the possible existence of axions has not been discarded  and has triggered a variety of experimental projects \cite{Cadene,PVLAS:2007wzd}.

Besides vacuum birefringence, the so-called {\it vacuum instability} refers to the possibility of production of electron-positron pairs excited when their rest mass energy threshold is available and arises in the theory  for values of magnetic/electric fields higher than the Schwinger critical  ones,  $B_c/E_c$, with  $E_c=\frac{m_e^2c^3}{e\hbar}=1.3\times10^{18}$ V/m, respectively.
Finally, {\it pressure anisotropy} associated to magnetized vacuum with a fixed background magnetic field, is another theoretical finding of QED in the one-loop approximation, somewhat explored in \cite{Elizabeth1,PPCF1}, where differences in parallel and transverse directions to the external field appear as a  consequence of the breaking of the SO(3) symmetry.

Apart from QED, there are additional alternative NLED that have been proposed in the literature incorporating quantum corrections. In the ModMax NLED \cite{ModMax}  SO(2) electromagnetic duality invariance and conformal invariance  are fulfilled. The Lagrangian density is not analytic everywhere failing at configurations for which the Lorentz invariants are zero. Other NLEDs, such as the Born-Infeld theory \cite{Borninfeld} instead, smoothes divergences and can be explored for strong EM fields as it gives the restriction on the
possible maximum electric field. Born-Infeld theory induces a dual invariance but displays no birefringence in vacuum \cite{nobirefrin}. Another difficulty is that the value of the electric field in the center of the point-like charge depends on the direction of approach to it but resolving this problem leads to application in gravity \cite{banados} and holographic superconductors.

The scope of this paper is revisiting the energy-momentum (EMT), angular momentum (AMT)
tensors and magnetic properties of the vacuum in selected NLED in a scenario where photons are propagating in the presence of a background magnetic field of arbitrary strength. We will restrict nevertheless to energy scales below the pair production instability i.e. photon frequency $\omega\leq 2 m_e c^2$.

In our study, we calculate the EMT opening the discussion related to the equivalence between the results from improved Noether and Hilbert EMT. All properties are obtained as  functions of derivatives of an effective Lagrangian with respect to the two scalar invariants of the theory, $\mathcal{F,G}$, thus becoming a more general study applicable to different NLEDs. In the same fashion,  all magnitudes obtained for the EH effective theory are valid for arbitrary orientation of the magnetic field, being equivalent cases that of a background pure external electric field or perpendicular electric and magnetic fields.

The paper is organized as follows. In Section \ref{section:waves} we introduce how the EM wave propagation can help study vacuum properties in the context of the NLED we consider. In section  \ref{section:1}  we present the effective NLED described by the Euler-Heisenberg Lagrangian for arbitrary magnetic field strength and study the propagation of a photon probe in its associated vacuum. We use a Lagrangian expansion in invariants of the theory along with coefficients, which are derivatives up to second order of it. This quadratic Lagrangian allows us to obtain generalized Maxwell equations and, in particular, that governing the displacement photon field and dispersion equations in the Fourier space.

In  section \ref{section:3} we proceed to obtain the EMT and AMT. Using the Hilbert construction, symmetric, gauge invariant, and conserved EMT is obtained. 
Additional physical quantities of the theory stemming from EMT and AMT are calculated, i.e. energy density, Poynting vector,  pressure components and angular momentum vector. We remark on differences found when using alternative procedures. In section \ref{sec:mag} we discuss the photon magnetization  and define a photon magnetic moment giving generic expression as a function of magnetic field strength in the range of validity of our approach and comparing it to existing limiting values. In section \ref{sec:exp} we follow with some discussion about possible strategies and experimental tests to discriminate among magnitudes obtained in the exposed scenarios. Finally, in \ref{conclusions}, our conclusions are presented. A final appendix section \ref{apendice} is included, where subsections detail complementary lengthy calculations appearing along this manuscript.

\section{Propagation of EM waves in vacuum}
\label{section:waves}
In order to further study vacuum properties as they arise in NLED theories special attention is devoted to wave propagation.
For example, radiation  (photon) emission from pulsars \cite{Mielniczuk,Taverna} i.e. magnetized neutron stars (NSs) has been reported to show the indirect signal of birefringence.  The latter has also been claimed responsible  in $e^+e^-$ production from photon fusion in Breit-Wheeler processes showing separation in the differential angular distribution relative to the initial photon polarization and magnetic field angle \cite{breit}.
The vacuum birefringence effect over the propagation of photons is  equivalent to that an ordinary anisotropic medium produces over light propagation. That means that vacuum {\it behaves} as a refractive medium. Hence, the study of propagation of photons in magnetized vacuum  using NLEDs in the lowest order of fields  allows us to interpret and design experimental tests for vacuum phenomenology using traditional treatments of non-linear optics. Only precise  observations and analysis of theoretical models may aim to isolate them. 
However, one should keep in mind that when these very large fields are generated in dense environments, matter effects are the main ingredient to include,  although  quantum vacuum aspects remain an important contribution indicating that a consistent treatment seems unavoidable.

The scenario depicted before is valid for field strengths smaller than the critical ones. However, generally speaking, magnetic field strengths in nature can reach extreme values, such as in the interior of white dwarfs or pulsars, where fields attain strengths $B\sim 10^{12-15}$ G \cite{Koe,Hard}, and those generated in heavy ion collisions. For the latter, and in the earliest moments after the collision, the system is subjected to what is expected to be the strongest magnetic field created in the laboratory. In heavy-ion collisions of $\mathrm{Au}-\mathrm{Au}$  at the RHIC energy $\sqrt{s_{N N}}=200\, \mathrm{GeV}$ and impact parameter $b=4\, \mathrm{fm}$, a local field $e B \approx 1.3 \cdot m_{\pi}^{2}$ with  $m_{\pi}^{2}=140^{2} \cdot 0.512 \times 10^{14}$ G $\approx 10^{18}$ G is estimated \cite{ionion,Blab}.

So far, there is a fundamental technical issue as it is not possible to generate steady fields stronger than $B\sim 4.5\times 10^{5}$ G in the lab because the magnetic stresses of such fields exceed the tensile strength of terrestrial materials. Regarding oscillating fields, special mention deserves those generated in laser facilities. For optical lasers, peak powers beyond 1 PW are available in several present and future projects such as in ELI\cite{eli}, CoReLS\cite{corels}, Appollon \cite{apollon}, Vulcan \cite{vulcan}  or CLPU  \cite{CLPU}, just to cite some of them. This would promote the laser intensity beyond that currently available at $\sim 10^{23}$ W cm$^{-2}$ \cite{corels}. For larger intensities close to $10^{24}$ W cm$^{-2}$ a 100 PW laser system is needed \cite{shen}. With better focusing, the laser intensity could be even higher. This will effectively drive laser-matter interaction in the strong-field QED regime. Note that for short times $\Delta t \ll P$, being $P$ the wave field period, and in localized regions, one can assume an equivalent constant magnetic field strength so the treatment remains valid. On the other hand in the context of Dirac materials \cite{diracmaterials} it is  possible to test the strong field properties of the vacuum of this theory with the advantage of the critical field being $\mathcal{O}(1)$ T.

In our treatment the NLEDs studied display magneto-electric properties in the magnetized vacuum. It is important to size at what extent they may affect the photon propagation. In this context, as in the effective theory of magnetized vacuum  \cite{Villalba-Chavez:2012pmx} Lorentz-symmetry is violated, the study of its properties could illuminate theories Beyond Standard Model that enable to  explain with non-conventional mechanisms, the existence of the so-called dark matter, milli-charged particles \cite{kouvaris} and other exotic phenomenology \cite{Cadene} that are, so far, out of the prediction of the standard scenarios.

\subsection{Effective EH Lagrangian for non linear electrodynamics}\label{section:1}

The one-loop photon-photon interaction processes are described by the  Euler-Heisenberg Lagrangian density, first proposed by Euler, Kockel \cite{HEKockel} and Heisenberg and independently by Weisskopf \cite{Weiss}. Following a renormalization procedure, it becomes finite and gauge invariant  \cite{ren,Schwinger} under the form
\begin{align}
&\mathcal{L}=-\mu_{0}^{-1} \mathcal{F}-\frac{1}{8 \pi^2}\int_{0}^{i\infty} \frac{d s}{s^{3}} e^{- m^2_e s } \nonumber\\
&\times\left [ (e s)^{2} a b \coth (e a s) \cot(e b s)-\frac{(e s)^{2}}{3}\left(a^{2}-b^{2}\right)-1
\right ],\label{eqEH}
\end{align}
where $a=\left[\left(\mathcal{F}^{2}+\mathcal{G}^{2}\right)^{1 / 2}+\mathcal{F}\right]^{1 / 2}$ and $b=\left[\left(\mathcal{F}^{2}+\mathcal{G}^{2}\right)^{1 / 2}-\right.$ $\mathcal{F}]^{1 / 2}$.
$m_{e}, e$ are the mass and charge of the electron, respectively. $\mathcal{F}, \mathcal{G}$ are secular invariants derived from the gauge and Lorentz invariants of the generic electromagnetic fields $({\bf E, B})$ defined as
\begin{align}
\mathcal{F}&=\frac{1}{4}F^{\mu\nu}F_{\mu\nu}=\frac{1}{2}(-\epsilon_0 E^2+\frac{B^2}{\mu_0}),\\ \mathcal{G}&=\frac{1}{4}F^{\mu\nu}\tilde{F^{\mu\nu}}=\sqrt{\frac{\epsilon_0}{\mu_0}}\boldsymbol{(-E.B)},
\end{align}

We have used $\mu_{0}$, $\epsilon_{0}$ as electrical permittivity and magnetic permeability of the empty space, respectively.  In what follows we set $\mu_{0}=\epsilon_{0}=1$ as well as the speed of light $c=1$ for the sake of simplicity.  The  field-strength tensor components are related to EM fields as $E_{i}=F_{0 i}$, $B_{i}=-\frac{1}{2} \epsilon_{i j k} F^{j k}$ with $i=1,2,3$. In addition, $\tilde{F}^{\mu\nu}=\epsilon^{\mu\nu\alpha\beta}F_{\alpha\beta}/2$ is the dual tensor and $\epsilon^{\mu\nu\alpha}$, $\epsilon^{\mu\nu\alpha\beta}$ are the totally antisymmetric Levi-Civita tensors of rank 3 and 4, respectively. We use the Einstein convention of summing over repeated indices. Note that for a pure magnetic field case (${\bf E}=0$) we have $\mathcal{F}=B^{2}/2$ and $\mathcal{G}=0$. Bianchi identities are fulfilled as $\frac{1}{\sqrt{-g}} \partial_{\nu} \left[ \sqrt{g} {\tilde F}^{\mu \nu}\right]=0$. At this point it is worth noting that for the sake of completeness we introduce the curved space notation, however, for most of the calculations we will restrict to a case assimilated to Minkowski flat space where we use the convention  $g_{\mu \nu}=\eta_{\mu\nu}=\rm{diag}(+1,-1,-1,-1)$  and $g$ for its determinant. 

In general, for arbitrary value of the EM fields or equivalently $a$, $b$ is not possible to get a handy  analytical expression of this Lagrangian in Eq. (\ref{eqEH}). Fortunately,  there is a more general treatment at $\mathcal{G}=0$ performing an expansion of the Lagrangian in terms of the Lorentz invariants. 
In particular, the scalar derivatives  $\mathcal{L}_{\mathcal{F}}=\frac{\partial \mathcal{L}}{\partial \mathcal{F}}$,
$\mathcal{L}_{\mathcal{FF}}=
\frac{\partial ^2\mathcal{L}}{\partial \mathcal{F}^2}$, $\mathcal{L}_{\mathcal{GG}}=\frac{\partial ^2 \mathcal{L}}{\partial\mathcal{G}^2}$ can be calculated analytically and we provide expressions in the limit $E\rightarrow0$, i.e. $b\rightarrow0$ which is the one we are interested in the present work, see  appendix \ref{apendice:1}.

In order to focus on the system under study we now introduce the general expression of the Euler-Heisenberg Lagrangian in Eq. (\ref{eqEH}) and decompose the EM field strength tensor as $\mathfrak{F}^{\mu \nu}(x) \equiv$ $F^{\mu \nu}(x)+f^{\mu \nu}(x)$ with contributions from the background (external) field $F^{\mu \nu}(x)=\partial^{\mu} A^{\nu}(x)-\partial^{\nu} A^{\mu}(x)$ and from the photon field $f^{\mu \nu}(x)=\partial^{\mu} a^{\nu}(x)-\partial^{\nu} a^{\mu}(x)$. From this we can construct the effective Lagrangian under the form

\begin{equation}
\mathcal{L}=-\frac{1}{4}\mathfrak{F}^{\mu \nu}\mathfrak{F}_{\mu \nu} +\mathcal{L}^{(1)} +\mathcal{L}^{(2)}.
\label{lagfF}
\end{equation}
The first term corresponds to the classical Maxwell Lagrangian and the last two terms correspond to quantum corrections up to second order in the radiation field strength tensor. $\mathcal{L}^{(1)}$ and  $\mathcal{L}^{(2)}$  describe interaction terms as \cite{Karbstein:2015cpa}

\begin{align}
\mathcal{L}^{(1)}&=\frac{1}{2}\mathcal{L}_{\mathcal{F}} f^{\mu\nu}F_{\mu\nu},\\
\mathcal{L}^{(2)}&=
\frac{1}{4}\mathcal{L}_{\mathcal{F}} f^{\mu\nu}f_{\mu\nu}\nonumber\\
&+\frac{1}{8}f^{\alpha\beta}\left ( \mathcal{L}_{\mathcal{F F}} F_{\alpha\beta}F_{\mu\nu}+ \mathcal{L}_{\mathcal{G G}} \tilde{F}_{\alpha\beta}\tilde{F}_{\mu\nu} \right )f^{\mu\nu}. \label{Lagra2}
\end{align}

The explicit expressions for the non-zero scalar derivatives $\mathcal{L}_{\mathcal{F}}=\frac{\partial\mathcal{L}}{\partial \mathcal{F}}|_{f=0}$,
$\mathcal{L}_{\mathcal{F F}}=\frac{\partial^2 \mathcal{L}}{\partial {\mathcal F}^2}|_{f=0}$ and $\mathcal{L}_{\mathcal{G G}}=\frac{\partial^2 \mathcal{L}}{\partial {\mathcal G }^2}|_{f=0}$ are shown in appendix  \ref{apendice:1} for completeness and have been previously compiled in \cite{DittrichLibro}.  $\mathcal{L}^{(1)}$ arises as a topological term in the expansion of Eq.(\ref{eqEH})  \cite{Karbstein:2015cpa,Bialynicka-Birula:2014bja}, { but giving null contribution to the equations of motion and the magnitudes arising from volume integrated space-time averaged quantities from EMT.} For the scope of this work we neglect higher orders  terms,  $\mathcal{L}^{(n)}$, $n>2$ which are indeed possible and can be generated in an analogous way. They would describe the  interaction of background and radiation fields as well as self-interaction to quartic order in the fields. From the perspective of  non-linear optical theory quartic terms in radiation field are related to possible new non-linear interactions, which do not occur at the tree level \cite{Battesti:2012hf} while for background fields is thus connected to the interaction of a magnetic field with electron-positron pairs capable to induce vacuum anisotropic pressures \cite{Elizabeth,PPCF1} similar to the ``Casimir effect''.

Therefore, the Lagrangian in Eq. (\ref{lagfF}) retains contributions from background external field (B) and photon  interaction with external fields (ph-B) as  $\mathcal{L}=\mathcal{L}^{\rm (B)}+\mathcal{L}^{\rm (ph-B)}$. The terms involving photons can be explicitly written as
\begin{align}
\mathcal{L}^{\rm (ph-B)}&=
-\frac{1}{4}(1-{\mathcal{L}}_{\mathcal F}) f^{\mu\nu}f_{\mu\nu}-
\frac{1}{2}(1-{\mathcal{L}}_{\mathcal F}) f^{\mu\nu}F_{\mu\nu}\nonumber\\
&+\frac{\mathcal{L}_{\mathcal{FF}}}{8} (f^{\mu\nu}F_{\mu\nu})^2
+ \frac{\mathcal{L}_{\mathcal{GG}}}{8} (f^{\mu\nu}\tilde{F}_{\mu\nu})^2.
\label{Lagrafoton}
\end{align}
The interaction terms take the following explicit form in terms of the photon  field $({\bf E}_w$, ${\bf B}_w)$ up to second order as well as the external magnetic field ${\bf B}_e\equiv {\bf B}$,

%with the additional parametrization  $\bf {B}_{tot}=B+B_w$ and $\bf{E}=E_w$ Lagrangian reads as

\begin{align}
\label{2orderLagrangianEB}
\mathcal{L}^{\rm (ph-B)}&=\frac{(1-\mathcal{L}_{\mathcal{F}})}{2}(E_w^2-B_w^2)+(1-\mathcal{L}_{\mathcal{F}})({\bf B}\cdot {\bf B}_w)\nonumber\\
&+ \frac{\mathcal{L}_{\mathcal{FF}}}{2}({\bf B} \cdot{\bf B}_w)^2+
\frac{\mathcal{L}_{\mathcal{GG}}}{2}({\bf B}\cdot {\bf E}_w)^2.
\end{align}
Note that this quadratic approximation for $\mathcal{L}^{\rm (ph-B)}$  has also been the starting point of previous studies \cite{DiPiazza:2002ve,Neves:2021tbt,Shabad:2011hf} in weak or strong field limits. It is important to recall that by using the soft-photon approximation in the Lagrangian in Eq.(\ref{2orderLagrangianEB}) we restrict the validity of our approach to a maximum value $B\leq 430 B_{c}$ \cite{DittrichLibro}.
Above this strength, we have to include two-loop corrections \cite{Villalba-Chavez:2009ruk} with explicit quantum treatment \cite{Chaichian,perez1,perez2,perez3}. However, for most usual  physical motivations: a study of laboratory astrophysics with pulsating laser fields and/or Neutron Star physics, a single loop approximation suffices, because typical magnetic fields in external layers and magnetospheres usually stay hundreds of times lower than the critical magnetic field.

Although the EH NLED  description may conceal some microscopic phenomena related to photon-photon interaction, it is a robust theory since it considers the vacuum behaves as a non-linear refractive optical medium.
As an example, from Eq. (\ref{2orderLagrangianEB}) and Lagrangian derivatives in appendix \ref{apendice:1} we recover the weak field (WF) approximation expressions $\mathcal {L}_{\mathcal {F}}=\frac{\xi\mathcal {F}}{4}$, $\mathcal {L}_{\mathcal {FF}}=\frac{\xi}{2}$, $\mathcal {L}_{\mathcal {GG}}=\frac{7\xi}{4}$ 
obtaining the well-known form
\begin{equation}
\mathcal{L}_{EH}^{WF}=-\mathcal{F}+\frac{\xi}{4}(4\mathcal{F}^2 + 7\mathcal{G}^2),   \label{nolinealL}
\end{equation}
usually taken for studies of non-linear laser optics \cite{Rizzo:2010di,Ferrando:2007pgk,Baier:2018vso}. 

From the Lagrangian $\mathcal{L}^{(\rm ph-B)}$ in Eq. (\ref{2orderLagrangianEB}) we can readily obtain Maxwell equations  (see appendix \ref{apendice:0}) and related  electric permittivity and magnetic permeability tensors.

%%%%%%%%%%%%%%%%%%%%%%%%%%%%%%%%%%%%%%%%%%%%%%%%
\begin{figure}[t!]
\centering
  \includegraphics[width=0.8\linewidth]{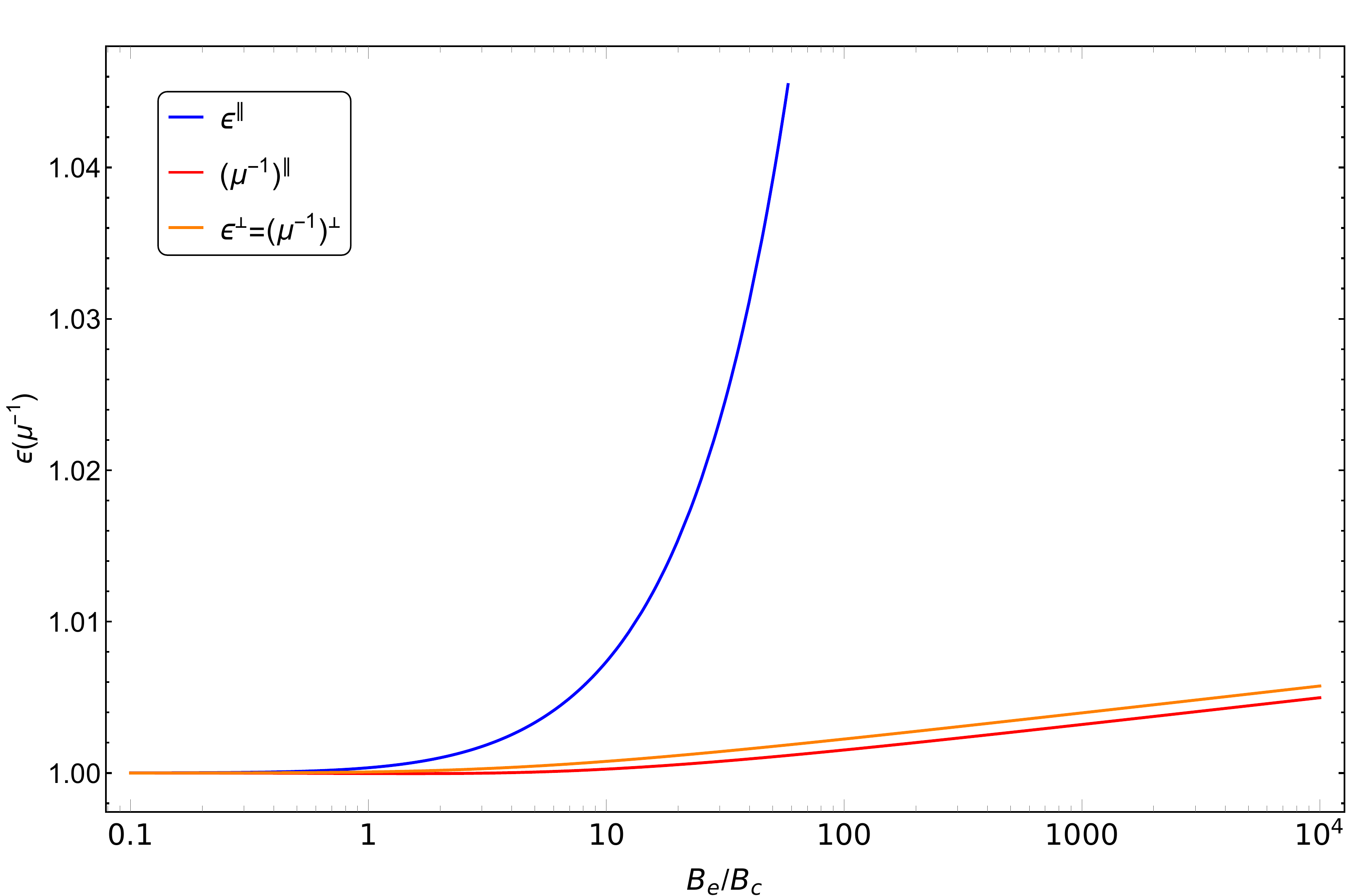}
 \caption{Electric permittivities and magnetic permeabilities versus magnetic field strength. These quantities remain positive definite ensuring the unitarity and causality constraint in vacuum (see appendix \ref{apendice:0}).}
 \label{fig:velocity}
\end{figure}
%%%%%%%%%%%%%%%%%%%%%%%%%%%%%%%%%%%%%

In Fig. (\ref{fig:velocity}) we have plotted electric permittivities and magnetic permeabilities from Eq. (\ref{epsilon}) and Eq. (\ref{epsimu2}) in appendix \ref{apendice:0} to illustrate these B-dependent quantities remain  positive definite as causality and unitarity require\cite{Shabad:2011hf},   $\mathcal{L}_{\mathcal{FF}} \geq 0$, $\mathcal{L}_{\mathcal{GG}}\geq 0$ and $1-\mathcal{L}_{\mathcal{F}}\geq 0$.

\section{Energy-momentum and Angular momentum tensors}
\label{section:3}

In this section, we are interested in determining the EMT and AMT, starting off from the Lagrangian in  Eq.(\ref{lagfF}) describing a photon probe propagating transverse to an external magnetic field.
There are various techniques at hand for this, stemming from the  effective theory and, usually, the approach chosen is based on its intended use.

Among the most common are those from Noether \cite{Landau}, based on the symmetries in the Lagrangian and  Noether theorem \cite{Noether} (current and charges are preserved), and Hilbert \cite{Hilbert}, which is based on variational geometry. Other several approaches have been developed to analyze the wave propagation problem, in particular those due to Boillat \cite{boillat}, Bialynicka-Birula and Bialynicki-Birula \cite{bb} and more recently by Novello et al. \cite{novello}. It is important to note that, in general, they do not yield the same result when computed. Just to mention some important differences, the Noether canonical procedure needs the Belinfante tensor to yield a symmetric tensor \cite{Belinfante}. Often the procedures providing gauge invariant, symmetric, and conserved EMT \cite{Blaschke:2016ohs,Baker:2020eqs} are best suited for studies with non-flat geometry such as those in General Relativity while Noether and its (improved) extensions for most of remaining  physical applications.

Despite the fact that we consider as a case study the EH  NLED whose limit at zero external magnetic fields is Maxwell electrodynamics, we will show in what follows that, as a clear example of the previous,  Noether and Hilbert approaches do not yield the same EMT tensor.

In the robust Hilbert method, we compute the EMT, $T_{H}^{\mu\nu}$, from the variation of the effective Lagrangian  $\mathcal{L}$ with respect to the metric tensor, $g^{\mu\nu}$ in the usual way. It has two main contributions
\begin{equation}
T_H^{\gamma\rho} =T_H^{(0) \gamma\rho} + t_H^{\gamma\rho},
%T_H^{(1) \gamma\rho} +T_H^{(2) \gamma\rho},
\end{equation}
where $T_H^{(0) \gamma\rho}$ is that of the external background field while $t_{H}^{\gamma\rho}$ is that accounting for the interaction of the photon field with the background field.
Therefore from the photon Lagrangian $\mathcal{L}^{\rm (ph-B)}$ 
\begin{equation}
t_{H}^{\gamma\rho} =\left.\frac{2}{\sqrt{-g}} \frac{\delta [\mathcal{L}^{\rm (ph-B)}]}{\delta g_{\gamma \rho}}\right|_{g=\eta}.
\label{metricder}
\end{equation}

Performing the derivatives and recovering the flat space

\begin{align}
t_H^{\gamma\rho}&=(1-\mathcal{L}_{\mathcal{F}}) f^{\gamma}_{\lambda}f^{\lambda \rho}+\frac{\mathcal{L}_{\mathcal{FF}}}{2} f^{\mu\nu}F_{\mu\nu}(F^{\gamma\alpha}f_{\alpha}^{\rho}+F^{\rho\alpha}f_{\alpha}^{\gamma})\nonumber\\
&+\frac{\mathcal{L}_{\mathcal{GG}}} {2} f^{\mu\nu} \tilde{F}_{\mu\nu}(\tilde{F}^{\gamma\alpha}f_{\alpha}^{\rho}
+\tilde{F}^{\rho\alpha}f_{\alpha}^{\gamma})+\frac{\eta^{\gamma\rho}}{4}\left((1-\mathcal{L}_{\mathcal{F}})
f_{\mu\nu}f^{\mu\nu}\right.\nonumber\\
&+\frac{\mathcal{L}_{\mathcal{FF}}}{2}f^{\mu\nu}F_{\mu\nu}f^{\alpha\beta}F_{\alpha\beta}+\frac{\mathcal{L}_{\mathcal{GG}}} {2}\left. f^{\mu\nu} \tilde{F}_{\mu\nu}f^{\alpha\beta}\tilde{F}_{\alpha\beta}\right)\nonumber\\
&+\mathcal{L}_{\mathcal{F}} (F^{\gamma\alpha}f_{\alpha}^{\rho}+F^{\rho\alpha}f_{\alpha}^{\gamma})+\frac{\eta^{\gamma\rho}}{2}\mathcal{L}_{\mathcal{F}}f^{\mu\nu} F_{\mu\nu}.\label{defEMT1}
\end{align}

\noindent The obtained EMT is thus gauge invariant, symmetric $t_H^{\mu\nu}=t_H^{\nu\mu}$,
displays  anisotropy and lack of tracelessness i.e.
it has a non-vanishing trace $t^{\mu}_{H,\mu}\neq 0$.

As we will check below, the energy momentum tensor $t_{H}^{\mu\nu}$ is locally conserved on shell imposing the corresponding equations of motion \cite{Blaschke:2016ohs}.

%%%%%%
 The photon EMT  can be also obtained as in  \cite{Neves:2021tbt,Villalba-Chavez:2012pmx}, using the improved Noether procedure yielding $t_{N+B}^{\mu \nu}$. We label this tensor as $N+B$ from Noether plus Belinfante.  Recapping the procedure we have

\begin{equation}
t_{N+B}^{\mu \nu}=2 \frac{\partial \mathcal{L}^{\rm(ph-B)}}{\partial f_{\mu \alpha}} f_\alpha^\nu-\eta^{\mu \nu} \mathcal{L}^{\rm(ph-B)},
\label{der}
\end{equation}

\noindent and its explicit expression is

%\begin{equation}
\begin{align}
t^{\mu\nu}_{N+B}&=
(1-\mathcal{L}_{\mathcal{F}})f^{\alpha\mu} f^{\nu}_{\alpha}+\frac{\mathcal{L}_{\mathcal{FF}}} {2} f^{\rho\sigma} F_{\rho\sigma} (F^{\mu\alpha}f_{\alpha}^{\nu})\nonumber\\
&+\frac{\mathcal{L}_{\mathcal{GG}}} {2} f^{\rho\sigma} \tilde{F}_{\rho\sigma} (\tilde{F}^{\mu\alpha}f_{\alpha}^{\nu})
-\eta^{\mu\nu}\mathcal{L}^{\rm (ph-B)}.
\label{noether}
\end{align}
%\end{equation}

The tensor in Eq. (\ref{noether})  is also gauge invariant and conserved but it is not symmetric and, therefore, is quantitatively different to Eq. (\ref {defEMT1}). The authors in \cite{Villalba-Chavez:2012pmx} claim the latter is due to the external magnetic field breaking the rotational symmetry thus not being possible to symmetrize $t_{N+B}^{\mu\nu}$. 

As is well known, both methods are mathematically quite different. In the Hilbert formalism, see Eq.(\ref{metricder}), it implies the variation of the Lagrangian with respect to the metric while for Noether, see Eq.(\ref{der}), it is varied with respect to the fields.

%Focusing in the model under study, 
The EMT in Eq. (\ref{noether}) and Eq. (\ref{defEMT1}), have been obtained starting from an approximate version of EH
Lagrangian, Eq.(\ref{Lagrafoton})
%, which is an expansion up to second order in the photon field. 
by performing an expansion including non-linear magneto-electric terms.
%of fourth order in the external magnetic field and photon fields.  
As is explained in \cite{Baker:2020eqs} the term proportional to $\eta^{\mu\nu}\mathcal{L}$ in the Noether EMT is recovered by the Hilbert method differentiating the factor $\sqrt{-g}$, hence, the differences come from the 
derivatives in the non-linear terms.
By Hilbert  method a term with a four metric tensor product emerges contributing to symmetrize the EMT while the corresponding one obtained by Noether method only yields one of these two terms and the resulting EMT is non-symmetric.
In theories like Classical Electrodynamics, identical results are obtained from Hilbert or Noether methods.
Thus one recovers Maxwell theory  from the EMT of a photon probe in a non-linear theory computed by Hilbert or Noether plus Belinfante symmetrization formalism, in the limit of $B_e\rightarrow 0$, being the tensor gauge invariant, symmetric, conserved and traceless.

\subsection{Equivalence of the EMT \`a la Noether and \`a la Hilbert for EH non linear electrodynamics }

At this point we aim to discuss the equivalence of the Noether or Hilbert calculation of the photon EMT if, instead of starting off from the approximate Lagrangian Eq. (\ref{Lagrafoton}), we use a  general Lagrangian $\mathcal{L}=\mathcal{L}(\mathcal{F},\mathcal{G})$ with $\mathcal{F}_{\mu\nu}={F}_{\mu\nu}+{f}_{\mu\nu}$ in the limit  ${f}_{\mu\nu}\ll {F}_{\mu\nu}$.
Thus from a general Lagrangian $\mathcal{L}$ following the prescribed procedure an equivalent EMT is obtained either by Hilbert or Noether plus Belinfante method.

The expansion up to second order in the photon fields of the general EMT obtained from $\mathcal{L}$ by the Hilbert method, also yields the EMT in Eq. (\ref{defEMT1}). Hence, with this method no matter if we compute the EMT from Lagrangian $\mathcal{L}^{\rm (ph+B)}$ or if we start from a general Lagrangian $\mathcal{L}$ provided we expand and retain terms up to second order in the photon fields. On the contrary, with the Noether method we can obtain the expression Eq. (\ref{defEMT1}) only by expanding up to second order in  the photon fields from a symmetric EMT obtained from  $\mathcal{L}$. 

\noindent Let us now illustrate the calculation   to clarify these ideas.  
\noindent We start from a general Lagrangian 
\begin{equation}
\mathcal{L}=\mathcal{L}(\mathcal{F,G})
\end{equation}
\noindent the  Noether plus Belinfante EMT $T_{N+B}^{\mu\nu}$ takes the form (note we consider the full EMT here)
\begin{align}
T_{N+B}^{\mu\nu}&=2\frac{\partial\mathcal{L}}{\partial\mathcal{F}_{\mu\alpha}}\mathcal {F}_{\alpha}^{\nu}-\eta^{\mu\nu}\mathcal{L}, \label{defT}
\end{align}
\noindent with
\begin{align}
\frac{\partial{\mathcal L}}{\partial\mathcal{F}_{\mu\alpha}}&=\frac{1}{2}\frac{\partial{\mathcal L}}{\partial\mathcal{F}}\mathcal{F}^{\mu\alpha} +\frac{1}{2}\frac{\partial{\mathcal L}}{\partial\mathcal{G}}\tilde{\mathcal{F}}^{\mu\alpha}, \label{derFG}
\end{align}

\noindent we obtain the symmetric tensor
\begin{equation}
T_{N+B}^{\mu\nu}=\frac{1}{2}\mathcal{L}_{\mathcal{F}}\mathcal{F}^{\mu\alpha}\mathcal {F}_{\alpha}^{\nu}+\frac{1}{2}\mathcal{L}_{\mathcal{G}}\tilde{\mathcal{F}}^{\mu\alpha}\mathcal {F}_{\alpha}^{\nu}-\eta^{\mu\nu}\mathcal{L}.\label{LFsimetrico}
\end{equation}

 Substituting in Eq.(\ref{LFsimetrico}) $(i)$ the expansion of $\mathcal{L}$ up to second order terms in the photon field as given by  Eq. (\ref{Lagrafoton}) and $(ii)$ the expansion of Lagrangian derivatives evaluated for $\mathcal{L}_{\mathcal{G} }=0$ and $\mathcal{L}_{\mathcal{FG}}=0$ one arrives at the same expression of $T_{N+B}^{\mu\nu}$ found in the Hilbert formalism. For this one must retain only up to second order terms on both the photon and the external constant fields and neglect terms proportional to $f^{\sigma\beta}f_{\sigma\beta}$, $\tilde{f}^{\sigma\beta}f_{\sigma\beta}$.   % Eq.(\ref{defEMT1}) from the approximate Lagrangian $\mathcal{L}^{\rm (ph-B)}$.

Let us now discuss the local conservation of the EMT Eq. (\ref{defEMT1}). Since we know
$T_{N+B}^{\mu\nu}=T_{H}^{\mu\nu}=T_{H}^{(0)\mu\nu}+t_{H}^{\mu\nu}$ we obtain that

\begin{align}
\partial_{\mu}t_{H}^{\mu\nu}
&=\frac{1}{2}(1-\mathcal{L}_\mathcal{F}) \left ((F^{\mu\nu}+f^{\mu\nu})\partial _{\mu}f_{\alpha}^{\nu}-(F^{\sigma\rho}+f^{\sigma\rho})\right )
\partial^{\nu}f_{\sigma\rho}
\nonumber\\
&+\frac{1}{2}\mathcal{L}_{\mathcal{FF}}F^{\alpha\beta}
f_{\alpha\beta}
(F^{\mu\alpha}\partial_{\mu}f_{\alpha}^{\nu} -F^{\sigma\rho}\partial^{\nu}f_{\sigma\rho})\nonumber\\
&+\frac{1}{2}\mathcal{L}_{\mathcal{GG}}\tilde{F}^{\alpha\beta}f_{\alpha\beta}(\tilde{F}^{\mu\alpha}\partial_{\mu}f_{\alpha}^{\nu}-\tilde{F}^{\sigma\rho}\partial^{\nu}f_{\sigma\rho}), \label{conservT}
\end{align}

\noindent where we have evaluated the divergence of the EMT on  the approximated  equation of motion
\begin{align}
\partial_{\mu}(\frac{\partial\mathcal{L}}{\partial\mathcal{F}_{\mu\nu}})=-\frac{1}{2}(1-\mathcal{L}_\mathcal{F})\partial_{\mu}f^{\mu\nu} +\partial_{\mu}\frac{1}{4}\mathcal{I}^{\mu\nu}=0,\label{EqM}
\end{align}

with
\begin{equation}
\mathcal{I}^{\mu\nu}=\left (\frac{\mathcal{L}_{\mathcal{FF}}}{2}(F^{\sigma\rho}f_{\sigma\rho}(F^{\mu\nu}+f^{\mu\nu}))
 +\frac{\mathcal{L}_{\mathcal{GG}}} {2}(\tilde{F}^{\sigma\rho}f_{\sigma\rho}(\tilde{F}^{\mu\nu}+\tilde{f}^{\mu\nu})) \right).\label{EqM1}
\end{equation}

\noindent Doing some algebra we can accommodate Eq.(\ref{conservT}) as

\begin{align}
\partial_{\mu}t_{H}^{\mu\nu}=\frac{1}{2}\mathcal{L}_{\mathcal{FF}}F^{\alpha\beta}f_{\alpha}^{\nu}\frac{1}{2}\eta^{\nu\lambda}F^{\mu\alpha}(\partial_{\mu}f_{\lambda\alpha}+\partial_{\lambda}f_{\alpha\mu}+\partial_{\alpha}f_{\mu\lambda})\nonumber\\
+\frac{1}{2}\mathcal{L}_{\mathcal{GG}}\tilde{F}^{\alpha\beta}f_{\alpha\beta}\frac{1}{2}\eta^{\nu\lambda}\tilde{F}^{\mu\alpha}(\partial_{\mu}f_{\lambda\alpha}+\partial_{\lambda}f_{\alpha\mu}+\partial_{\alpha}f_{\mu\lambda})\nonumber\\
\frac{1}{2}(1-\mathcal{L}_\mathcal{F})\frac{1}{2}\eta^{\nu\lambda}\tilde{F}^{\mu\alpha}(\partial_{\mu}f_{\lambda\alpha}+\partial_{\lambda}f_{\alpha\mu}+\partial_{\alpha}f_{\mu\lambda})=0,
\end{align}

\noindent due to the Bianchi identity $(\partial_{\mu}f_{\lambda\alpha}+\partial_{\lambda}f_{\alpha\mu}+\partial_{\alpha}f_{\mu\lambda})=0.$

As we have checked, the second order on photon field symmetric EMT is conserved.
Let us note that the equation of motion,  Eq. (\ref{EqM}) fulfills
, contains quadratic terms in the photon fields
\begin{equation}
\frac{\mathcal{L}_{\mathcal{FF}}}{2}F^{\sigma\rho}f_{\sigma\rho}f^{\mu\nu}+\frac{\mathcal{L}_{\mathcal{GG}}} {2}\tilde{F}^{\sigma\rho}f_{\sigma\rho}\tilde{f}^{\mu\nu}=0,
\end{equation}

\noindent 
and after neglecting them 
and we obtain a similar expression in accordance to the one obtained by \cite{Shabad:2011hf} for $\mathcal{L}^{\rm (ph-B)}$.

In our previous work \cite{Perez-Garcia:2022rzy},  the  Hilbert-Einstein  method was also used to obtain the diagonal components of the EMT in the weak field limit in Eq. (\ref{nolinealL}). That calculation was performed considering the external plus wave fields contributions to the strength  tensor i.e. $F^{\mu\nu} + f^{\mu\nu}$ but neither were explicitly separated nor the latter was approximated to their quadratic contributions. Hence, it is worth noting that the resulting  EMT contains all possible interactions between the external magnetic and the photon  field probes, including  self-interaction, that, nevertheless, should coincide with the $t_H^{\mu\nu}$ presented here if we make the appropriate approximations above mentioned. This  result is not a priori straightforward to obtain \cite{Noble,Dereli et al.(2007)}.
%%%%%%%%

\textbf{Going back to the  previous disscusion about Noether and Hilbert formalisms using ${\mathcal L}^{\rm (ph-B)}$,  as both produce  different but conserved EMT (regarding improved Noether the continuity equation is limited to second index) in principle they should be both acceptable. The physical meaning of the two  approaches  and the implications of their  differences for  the observables is still an open question and deserves further work to be fully understood, being  however beyond the scope of this work. Nevertheless in the next two subsection we analyze some properties and characteristics of  Noether and Hilbert photon EMTs in  Eqs.(\ref{defEMT1}) and (\ref{noether}).}
%One possibility is that the differences will be determined by consider different setups.}

\subsection{Components of $t_{H}^{\mu\nu}$}
Let us continue discussing the explicit components of the photon contribution $t_{H}^{\mu\nu}$ obtained from Eq.(\ref{defEMT1}) and doing comparison with the result obtained by non-symmetric Noether improved method Eq(\ref{noether}).  We can write the components in terms of the photon fields ${\bf H}_{w}$,${\bf B}_w$, ${\bf D}_w$ and ${\bf E}_w$, see Eq. (\ref{dw}) and Eq.(\ref{mag}) as
 \begin{align}
  t_H^{00}&=\frac{1}{2} (\mathbf{D}_w.\mathbf{E}_w+
 \mathbf{H}_w.\mathbf{B}_w)+\mathcal{L}_{\mathcal{GG}}(\mathbf{B}_{e}\cdot\mathbf{E}_w)^2,
 %\frac{3}{2} (\mathbf{D}_w.\mathbf{E}_w+
%\mathbf{H}_w.\mathbf{B}_w), \label{energy0}
\end{align}
 while components  $i=j=1,2$ are given by
\begin{align}
 t^{ii}_H&=\frac{1}{2}(\mathbf{D}_w.\mathbf{E}_w+\mathbf{H}_w.\mathbf{B}_w)\nonumber\\
& -(D_{w,i}E_{w,i} +H_{w,i}B_{w,i})-\mathcal{L}_{\mathcal{FF}}(\mathbf{B}_{e}\cdot\mathbf{B}_w)^2.
\label{EMT-SWfield1}
\end{align}

\noindent Instead, for $i=j=3$,  we obtain
\begin{align}
 t_H^{33}&=\frac{1}{2}(\mathbf{D}_w.\mathbf{E}_w+\mathbf{H}_w.\mathbf{B}_w)\nonumber\\
  &-(D_{w,3}E_{w,3} +H_{w,3}B_{w,3})-\mathcal{L}_{\mathcal{GG}}
(\mathbf{B}_{e}\cdot\mathbf{E}_w)^2,
\label{EMT-SWfield2}
\end{align}
\noindent while for the ${\bf B_e}=B_e{{\bf \hat z}}$ the only spatial non-diagonal component $i\ne j$ is

\begin{align}
t_H^{13}=t_H^{31}=&-\mathcal{L}_{\mathcal{GG}}
(\mathbf{B}_{e}\cdot\mathbf{E}_w)E_{w,1}B_{e,3} \nonumber\\
&+\mathcal{L}_{\mathcal{FF}}
(\mathbf{B}_{e}\cdot\mathbf{B}_w)B_{w,1}B_{e,3}=0,
\end{align}
and the non-vanishing components $0i$ corresponding to the momentum density, $\pmb{ \mathcal{P}}_w$, are
 \begin{align}
&{t}_H^{02}=
{\frac{1}{2}}\left(
(E_{w,3}H_{w,1}-E_{w,1}H_{w,3})+(D_{w,3}B_{w,1}-D_{w,1}B_{w,3})\right. \nonumber\\
&+{\mathcal L}_{\mathcal{FF}} E_{w,1} B_{e,3} (\mathbf{B_e}\cdot {\bf B}_w)
+{\mathcal L}_{\mathcal{GG}} \left.B_{w,1} B_{e,3}({\bf B_e}\cdot {\bf E}_{w})\right).
\label{poyntingw}
\end{align}

\noindent  Besides, the AMT, a rank-3 tensor density, connected to the symmetric EMT via tensor relations \cite{Blaschke:2016ohs}  as
%
%To complete the understanding of the non-linear theory  we must
%also draw our attention to
\begin{align}
\mathfrak{M}^{\mu\nu\lambda}&= x^{\mu}t_H^{\nu\lambda}(x)-x^{\nu}t_H^{\mu \lambda}(x),
\label{cuatritorque1}
\end{align}
%begin{align}
%\label{cuatritorque2}
% \end{align}

%\noindent noting that it is conserved $\partial_{\lambda}\mathfrak{M}^{\mu%%%\nu\lambda} =T_H^{\nu\mu}-T_H^{\mu \nu}=0,$ and antisymmetric in the $\mu \nu$ indices.

\noindent is conserved $\partial_{\lambda}\mathfrak{M}^{\mu\nu\lambda} =t_H^{\nu\mu}-t_H^{\mu \nu}=0,$
and antisymmetric in the $\mu \nu$ indices.

Once we have defined the tensorial quantities of interest, EMT and AMT, as space-time dependent we now proceed to provide their physical meaning. To do that we express the explicit components of the photon contribution to  $t_H^{\mu\nu}$ in terms of the fields  ${\bf B}={\bf B}_e+{\bf B}_w$, ${\bf E}={\bf E}_w$,  obtaining magnitudes such as angular momentum, energy density, Poynting vector or directional pressures.

On general grounds, magnitudes being conserved stem from the dynamical equations they follow in a system. Then, EMT and AMT conservation laws usually lead to the standard procedure to define  physical quantities  as integrals of {\it densities} over a  spatial volume $V$ \cite{Villalba-Chavez:2012pmx,BirulaBook}. At the same time, the presence of oscillating fields leads to the  importance of averaging over time $t$ and coordinate propagation, $y$, whose  associated Fourier coordinates are frequency $\omega$ and momentum $\pmb{k}_\bot$, respectively.

In that way, the EH NLED vacuum would be described as a classical Maxwell anisotropic  medium and  the photon's propagation would proceed via an effective metric to fulfill the null geodesic requirement \cite{novello}.\\

\subsection{Angular momentum, energy density, pressures and Poynting vector}

 We now discuss the relevant  physical implications of Hilbert and Noether procedures. We first focus on the angular momentum vector $\pmb{\mathcal{J}}$ obtained from the AMT in Eq. (\ref{cuatritorque1}). As the obtained Hilbert EMT is symmetric its conservation is straightforward \cite{Blaschke:2016ohs}.   $\pmb{ \mathcal{J}}$ is given by volume integration as
\begin{align}
{\mathcal{J}}^k&=\frac{1}{2} \epsilon^{kij}\langle \mathfrak{M}^{ij0}\rangle_V =\frac{1}{2} \epsilon^{kij}\langle x^i t_H^{j0}-x^jt_H^{i0} \rangle_V, \label{torque}
\end{align}

\noindent with $i,j,k=1,2,3$. The mechanical relation between the EMT and AMT can be established \cite{Blaschke:2016ohs}. Hence, a conserved angular momentum vector  also leads to a zero torque,  $\frac{d {\pmb {\mathcal J}}}{d t}=\pmb{ \tau}=0$ and zero perpendicular component of the magnetization ($\pmb{ \tau}=\pmb{\mathcal M}_{\omega,\bot}\times{\bf B}_e$).

However, as shown in \cite{Villalba-Chavez:2012pmx} a non-conserved angular momentum vector, non-zero torque, and perpendicular magnetization arises  from a non-symmetric EMT, see Eq. (\ref{noether}), as when using the Noether plus Belinfante or Hamiltonian method. In such a case, an intrinsic photon spin could be defined \cite{Obukhov:2008dz}.

Let us now write the expression for the photon energy density $\rho_w=t_H^{00}$ and its integrated value, the photon energy contribution, ${\mathcal P}_w^{0}=\langle \rho_w \rangle_V$. The space-time averaged form, denoted with $\langle .. \rangle$, for $\mathcal{ P}_w^{0}$ is

%For this, let us consider ${\bf D}_{w}={\bf D}_{w\parallel} + {\bf D}_{w\bot}$ and ${\bf B}_{w}={\bf %B}_{w\parallel} + {\bf B}_{w\bot}$,
%using $E_{w,i}=\epsilon_{ij}^{-1} D_{w,j}$   where $\epsilon_{ij}^{-1}$ is the inverse dielectric %tensor and $H_{w,i}=\mu_{ij}^{-1} B_{w,j}$, see Eq.(\ref{epsilon}) to Eq.(\ref{epsimu2}).

\begin{align}
{\mathcal P}_{w,H}^{0}&=\left\langle \frac{D_{w}^2}{2\epsilon_{\bot}}+\frac{B_{w}^2}{2\mu_{\bot}}\right\rangle \nonumber\\
& +\frac{1}{2}\left\langle\frac{{ 3}\mathcal{L}_{\mathcal{GG}} ({\bf B}_e\cdot {\bf D}_w){\bf D}_w}{\epsilon_{\bot}\epsilon_{\parallel}}
-\mathcal{L}_{\mathcal{FF}}({\bf B}_e\cdot {\bf B}_w){\bf B}_w \right\rangle {\bf B}_e, \label{energy00}
\end{align}

 \begin{align}
{\mathcal P}_{w,N}^{0}&=\left\langle \frac{D_{w}^2}{2\epsilon_{\bot}}+\frac{B_{w}^2}{2\mu_{\bot}}\right\rangle \nonumber\\
 &+\frac{1}{2}\left\langle\frac{\mathcal{L}_{\mathcal{GG}} ({\bf B}_e\cdot {\bf D}_w){\bf D}_w}{\epsilon_{\bot}\epsilon_{\parallel}}
-\mathcal{L}_{\mathcal{FF}}({\bf B}_e\cdot {\bf B}_w){\bf B}_w \right\rangle {\bf B}_e, \label{energy00}
\end{align}

 We remind at this point that in general ${\mathcal P}_{w,H}^0$ is not the Hamiltonian $H$ because it does not fulfill the Legendre transformation
$$H={\bf D}_w \cdot {\bf E}_w({\bf D}_w,{\bf B}_w)-\mathcal{L}^{\rm (ph-B)}({\bf D}_w,{\bf B}_w).$$

Only for pure photon polarization mode (3)  ${\mathcal P}_w^0\equiv H.$  On the contrary, Noether $t_N^{00}$ always leads to the Hamiltonian \cite{Blaschke:2016ohs,Bialynicka-Birula:2014bja,Villalba-Chavez:2012pmx}. 

Regarding the Poynting vector,  $\pmb{{\mathcal P}}_{w,H}$,
we can obtain it from the EMT components $t_H^{0i}$. Due to the plane wave photon field, propagating in the $\pmb{\hat{y}}$-direction the only non-trivially fulfilled EMT conservation relation concerns the $i=2$ component. Having a fixed orientation in the $\pmb{\hat{z}}$-direction for the background field the only non-zero component is ${\mathcal P}_{w,H}^{2}=\langle t_H^{02}\rangle=\langle t_H^{20}\rangle $

\begin{align}
&\pmb{{\mathcal P}}_{w,H}=\langle D_{w,3}B_{w,1}\left(\frac{1}{\epsilon_{\parallel}\mu_{\bot}}+1\right)-D_{w,1}B_{w,3}
\left(\frac{1}{\epsilon_{\bot}\mu_{\parallel}}+1\right)\rangle \pmb{\hat y}\nonumber\\
&+\langle E_{w,1} B_{e,3}(\mathbf{B_e}\cdot \mathbf{B}_w){\mathcal L}_{\mathcal{FF}}+B_{w,1}B_{e,3}({\bf B_e}\cdot {\bf E}_{w}){\mathcal L}_{\mathcal{GG}}\rangle \pmb{\hat y}\label{Poynting},
\end{align}

\noindent and it does not match the one obtained by the Noether method,
\begin{align}
\pmb{\mathcal {P}}_{w,N}=\langle t_{N}^{02}\rangle &= \langle D_{w,3}B_{w,1}(\frac{1}{\epsilon_{\parallel}\mu_{\bot}}+1)\rangle\nonumber\\
&-\langle D_{w,1}B_{w,3}(\frac{1}{\epsilon_{\bot}\mu_{\parallel}}+1)\rangle \pmb{\hat y}.
\end{align}

\noindent and
\begin{align}
&\pmb{\mathcal {P}}_{w,N}^{momentum}=\langle t_{N}^{20}\rangle=\pmb{\mathcal {P}}_{w,N}\nonumber\\
&-\langle E_{w,1} B_{e,3}(\mathbf{B_e}\cdot \mathbf{B}_w){\mathcal L}_{\mathcal{FF}}\rangle+\langle B_{w,1}B_{e,3}({\bf B_e}\cdot {\bf E}_{w}){\mathcal L}_{\mathcal{GG}}\rangle \pmb{\hat y}.
\end{align}

\noindent Noether method yields an EMT such that  $t_{N}^{20}\equiv E_w\times H_{w}$ while
$t_{N}^{02}\equiv D_w\times B_{w}$ i.e.
$t_{N}^{20}\neq t_{N}^{02}$ due to the approximated  $\mathcal{L}^{\rm(ph-B)}$ used. 

As for  the photon pressure components, they are related to the stress energy tensor (spatial part in EMT)  $t_H^{ij}$. We thus consider the flux of the i-th component of momentum carried in the j-th direction and viceversa. As the background field points in the $\pmb{{\hat z}}$-direction from Eq. (\ref{EMT-SWfield1}) the expression for  $p_w^1=\langle t_H^{11}\rangle $, $p_w^2=\langle t_H^{22}\rangle$ reads

\begin{align}
p_{w,H}^{i}&=\langle \frac{1}{2}(\mathbf{D}_w\mathbf{E}_w+\mathbf{H}_w \mathbf{B}_w)\nonumber\\
&-(D_{w,i}E_{w,i} +H_{w,i}B_{w,i})-\mathcal{L}_{\mathcal{FF}}(\mathbf{B}_{e}\cdot\mathbf{B}_w)^2\rangle,
\label{EMT-SWfield11}
\end{align}
\noindent where $i=1,2$, while for $p_{w,H}^3=\langle t_{H}^{33}\rangle$ we obtain
\begin{align}
p_{w,H}^3&= \langle \frac{1}{2}(\mathbf{D}_w\mathbf{E}_w+\mathbf{H}_w\mathbf{B}_w)\nonumber\\
 &-(D_{w,3}E_{w,3} +H_{w,3}B_{w,3})-\mathcal{L}_{\mathcal{GG}}
(\mathbf{B}_{e}\cdot\mathbf{E}_w)^2\rangle.
\label{EMT-SWfield22}
\end{align}

The  expressions in Eq.(\ref{EMT-SWfield11}), (\ref{EMT-SWfield22}) clearly show the anisotropy regarding photon propagation with respect to the fixed direction of the magnetic field $\mathbf{B}_{e}$, breaking the rotational symmetry that otherwise would exist in  vacuum.
Instead, the calculation of Noether pressures  yield
\begin{equation}
p_{w,N}^{i} =\langle \frac{1}{2}(\mathbf{D}_w\mathbf{E}_w+\mathbf{H}_w\mathbf{B}_w) -(D_{w,i}E_{w,i} +H_{w,i}B_{w,i}) \rangle,
\label{EMT-presNoether}
\end{equation}
\noindent for $i=1,2,3$.  Although $p_{w,N}^{i}$ is also anisotropic it is different to those calculated in the Hilbert procedure.

Even when all pressures are different: $p_w^1$  $p_w^2$ and $p_w^3$, as the system becomes axially symmetric it seems natural to define non-linear pressures into parallel, $P_{w,\parallel}^{NL}$, and transverse, $P_{w,\bot}^{NL}$, directions to the background external field under the form

\begin{equation}
P_{w,\bot}^{NL}=\frac{ p^1_{w} + p^2_{w}}{2},\quad P_{w,\parallel}^{NL}= p^3_{w} .
\end{equation}
\noindent The non-linear (NL) contribution to photon energies ${\mathcal P}^{0,NL}_w \equiv E_w^{NL}$, Poynting vector ($\pmb{{\hat y}}$-direction component) and directional pressures using the Hilbert and Noether procedures for each of two physical polarization modes, see appendix \ref{apendice:0}, can be summarized (beware of notation for photon $E_w$ field and non-linear photon energy ${E}_{w}^{NL}$) as

\begin{align}
{E}_{w}^{H, NL,(2)}&={E}_{w}^{N,NL,(2)}+\mathcal{L}_{\mathcal{GG}}B_e^2\langle E_{w}^2\rangle\simeq (-\mathcal{L}_{\mathcal{F}}+{\frac{3}{2}}\mathcal{L}_{\mathcal{GG}}B_e^2)\langle E_{w}^2\rangle \\
{E}_w^{H,NL,(3)}&={E}_{w}^{N,NL,(3)}\simeq -(\mathcal{L}_{\mathcal{F}}+\frac{1}{2}\mathcal{L}_{\mathcal{FF}}B_e^2)\langle E_{w}^2\rangle,\nonumber \\
\mathcal{P}_w^{H,NL,(2)}&=\mathcal{P}_w^{N,NL,(2)}+\mathcal{L}_{\mathcal{GG}}B_e^2\langle E_{w}^2\rangle\simeq ( -\mathcal{L}_{\mathcal{F}}+\mathcal{L}_{\mathcal{GG}}B_e^2)\langle E_{w}^2\rangle\\
\mathcal{P}_w^{H,NL,(3)}&=\mathcal{P}_w^{N,NL,(3)}\simeq-(\mathcal{L}_{\mathcal{F}}+\mathcal{L}_{\mathcal{FF}}B_e^2)\langle E_{w}^2\rangle\nonumber \\
P^{H,NL, (2)}_{w,\parallel}&=P^{N, NL, (2)}_{w,\parallel}+\mathcal{L}_{\mathcal{GG}}B_e^2\langle {E}_{w}^2\rangle,\simeq-\frac{3}{2}\mathcal{L}_{\mathcal{GG}} B_e^2\langle E_w^2\rangle  \nonumber\\
P^{H,NL, (2)}_{w,\bot}&=P^{N,NL, (2)}_{w,\bot}\simeq-\frac{1}{2}(\mathcal{L}_{\mathcal{F}}-\mathcal{L}_{\mathcal{GG}}B_e^2)\langle E_{w}^2\rangle,\label{Pradiation2}\\
P_{w,\parallel}^{H,NL,(3)}&=P_{w,\parallel}^{N,NL,(3)}\simeq\frac{1}{2}\mathcal{L}_{\mathcal{FF}}B_e^2\langle E_{w}^2\rangle,\nonumber\\
P^{H,NL, (3)}_{w,\bot}&=P^{N,NL, (3)}_{w,\bot}+\mathcal{L}_{\mathcal{FF}}B_e^2\langle E_{w}^2\rangle\simeq -\frac{1}{2}(\mathcal{L}_{\mathcal{F}}
+3\mathcal{L}_{\mathcal{FF}}B_e^2)\langle E_{w}^2\rangle,\label{Pradiation3}
\end{align}

\noindent where we have  used $\frac{D_w^2}{\epsilon_{(\parallel,\bot)}}=\frac{ B_w^2}{\mu_{(\bot,\parallel)}}$ and $ B_w=E_w$.

\begin{figure}[t]
\centering
 \includegraphics[width=0.48\linewidth]{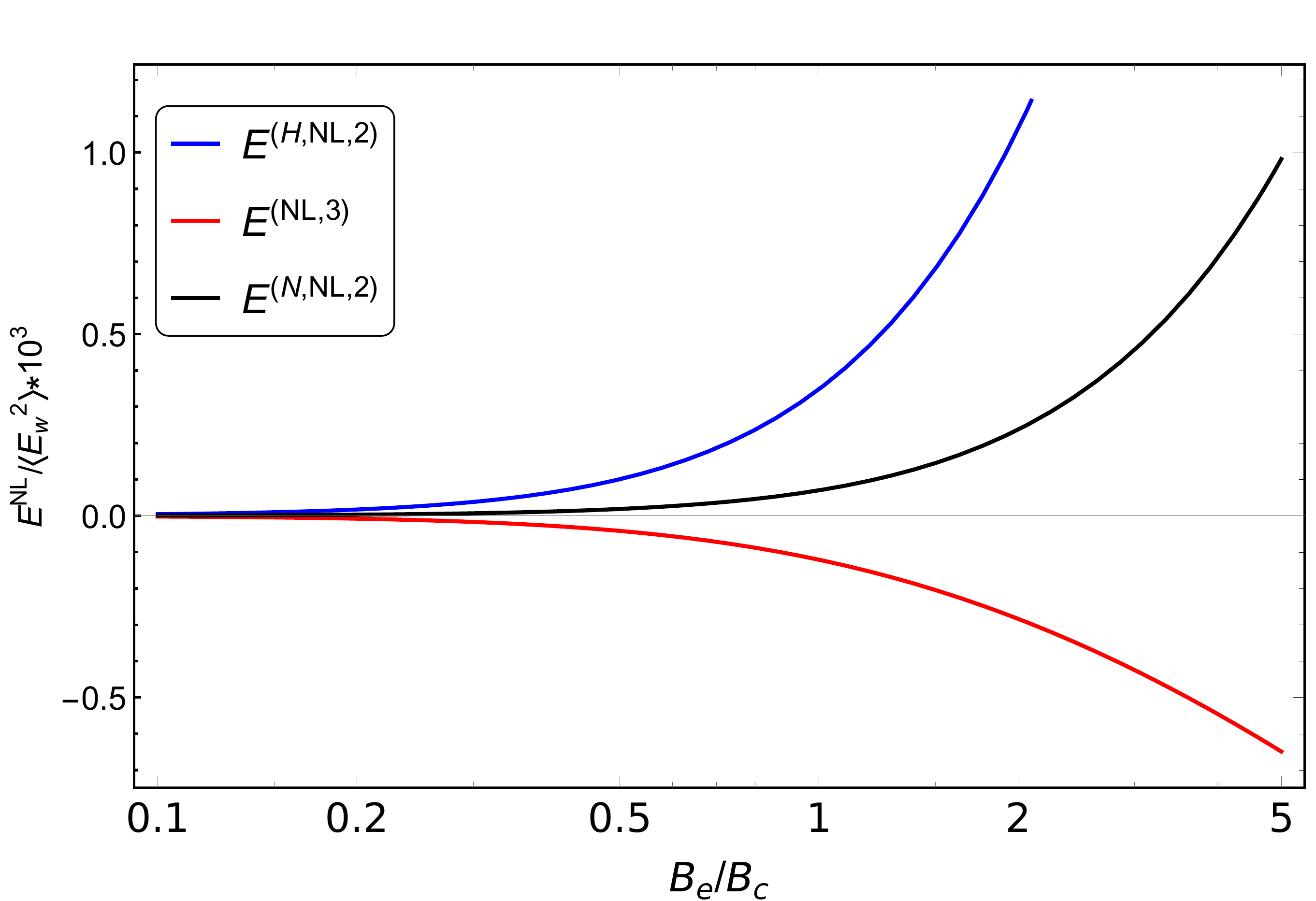}
\includegraphics[width=0.48\linewidth]{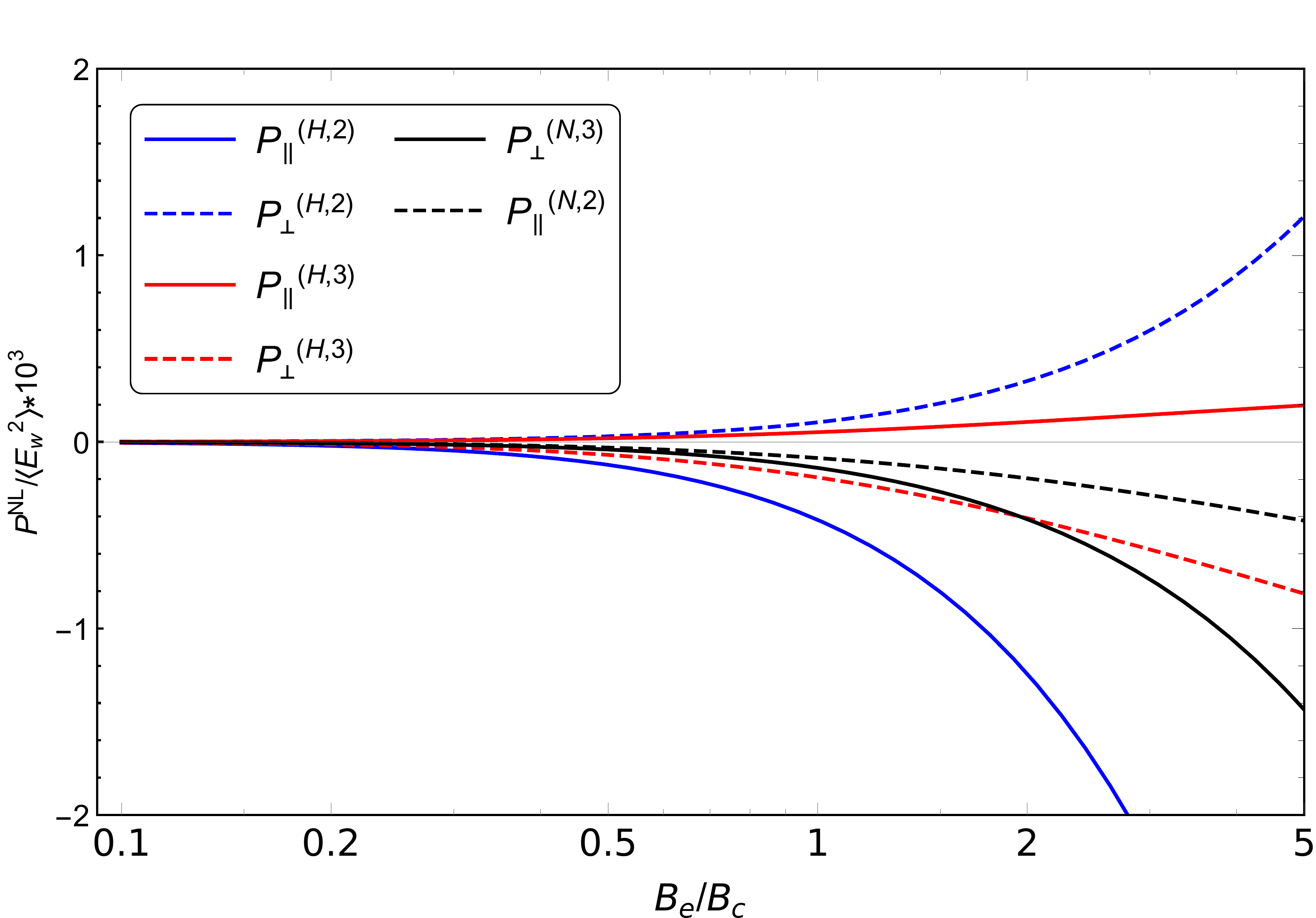}
 \caption{Non-linear contribution to energy $E^{NL}\equiv \langle \rho_w\rangle_V$ (left) and  parallel and perpendicular pressures (right) as a function of magnetic field strength for both polarization modes. ${E}^{NL,(2)}>{E}^{NL,(3)}$ with increasing magnetic field strength. ${E}^{NL,(3)}\equiv {E}^{N,NL,(3)}={E}^{H,NL,(3)}$. For mode (2),  $P^{NL,(2)}_{\bot}> P^{NL,(2)}_{\parallel}$ while $P^{NL,(3)}_{\bot}< P^{NL,(3)}_{\parallel}$, for mode (3).
}\label{fig:energies}
\end{figure}

Hilbert energy ${E}_{w}^{H,NL}$ has an extra term when compared to the Noether energy  for mode (2) while for mode (3) remains unchanged. The presence of magneto-electric terms, also in the magnetization is the main reason for this. Concerning pressures, we find that, depending on the photon polarization  mode,  differences appear in the parallel or perpendicular components. For polarization mode (2) perpendicular pressures are equal  while the parallel differs. On the contrary, for mode (3), the  perpendicular pressure differs in Hilbert and Noether's methods while the parallel are equal.
 In Fig. (\ref{fig:energies}) we have plotted  the non-linear contribution to photon energy ${E}_w^{NL}\equiv \langle \rho \rangle_V$ as obtained in Hilbert and Noether  procedures (left) and  parallel and perpendicular photon pressures (right) as a function of magnetic field strength for both polarization modes. For increasing magnetic field strength, { ${E}_w^{H,NL,(2)}>{E}_w^{H,NL,(3)}$} and $P^{NL,(2)}_{w,\bot}> P^{NL,(2)}_{w,\parallel}$ while for mode (3) it is the opposite,  $P^{NL,(3)}_{w,\bot}< P^{NL,(3)}_{w,\parallel}$ and ${E}_w^{H,NL,(3)}={E}_w^{N,NL,(3)}$.

%%%%%%%%%%%%%%%%%%%

Let us note that the photon pressure for monochromatic planar waves for perpendicular propagation of the photon, proceeds in $\pmb{\hat{y}}$-direction, defining $p_{w}^2$ the target pressure.
 Then, radiation pressure is defined by Eq (\ref{EMT-SWfield11}) being for  mode (2)
\begin{equation}\label{P2mode2}
p_{w}^2=(1-\mathcal{L}_{\mathcal{F}}+\frac{\mathcal{L}_{\mathcal{GG}}B^2}{2})\langle E_{w}^2\rangle,
\end{equation}

\noindent while for mode (3)
\begin{equation}\label{P2mode3}
 p_{w}^2=(1-\mathcal{L}_{\mathcal{F}}-\frac{3}{2}\mathcal{L}_{\mathcal{FF}}B^2)\langle E_{w}^2\rangle.
 \end{equation}

Although the Poynting vector is defined without ambiguity due the EMT symmetry, it does not fulfill that $p_w^2= \mathcal{P}_w$, as for photons  propagating in isotropic media or in  ``empty vacuum''.
This is due to external magnetic field breaking the rotational symmetry.

We note the  radiation pressure as well as the perpendicular pressure $P^{NL}_{w,\bot}$ have different behavior depending on their  propagation mode.
For mode (2) both are  higher  than the corresponding ``classical pressures'' while for mode (3) they become lower than the classical  value. At first glance the behavior of mode (2) seems to be counterintuitive. Although  photon propagating in magnetized vacuum has a lower velocity  than if it propagates in an ``empty vacuum'',  the contribution of magnetic energy enters in the radiation pressure as an additive term yielding a higher pressure than in the classical case, see Eq.(\ref{P2mode2}). Instead, for mode (3) this contribution appears subtracting, yielding thus a lower pressure value, see Eq.(\ref{P2mode3}).

\textbf{A more detailed analysis of the physical meaning of photon EMTs obtained from the Lagrangian $\mathcal{L}^{\rm (ph-B)}$ by Noether or Hilbert methods requires to go from the  local conservation to the calculation  of experimentally measurable physical quantities (like the radiation pressure). The experimental setup will then determine the boundary conditions to be imposed, both on the photon field and the background magnetic field. It is an open question to determine if those conditions will lead to different experimental results. Such an issue requires a deeper analysis that deserves a separate work in the future.}

\section{Magnetization}
\label{sec:mag}
%\subsubsection{Photon effective magnetic moment}
The magnetic energy density associated with the photon can be obtained from the magnetization in presence of an external magnetic field as,
\begin{equation}
\mathcal{E}_{mag}=-\frac{1}{2} \pmb{\mathcal{M}}_w \cdot\mathbf{B}_e
%= -\frac{1}{2} \left %%(\pmb{\mathcal{M}}_{w\,\parallel}\cdot {\bf B}_e \right )
%+ \mathcal{M}_{\bot}\cdot {\bf B}_w\right ).
\label{Emag}
\end{equation}

\noindent where the magnetization is calculated generically  from $\mathcal{L}^{(\rm ph-B)}$ in Eq. (\ref{Lagrafoton})  using ${\bf H}=-\frac{\partial \mathcal{L}^{(\rm ph-B)}}{\partial {\bf B}_e} = {\bf B}_e-\pmb{\mathcal{M}}$ or by the Hamiltonian $H^{(\rm ph-B)}$

\begin{align}
H^{(\rm ph-B)}&=\int d^3 x \frac{D_{w}^2}{2\epsilon_{\bot}}+\frac{B_{w}^2}{2\mu_{\bot}} \nonumber\\
&-\frac{1}{2}\left(  \frac{\mathcal{L}_{\mathcal{GG}}}{\epsilon_{\bot}\epsilon_{\parallel}} ({\bf B}_e\cdot {\bf D}_w)^2
-\mathcal{L}_{\mathcal{FF}}({\bf B}_e\cdot {\bf B}_w)^2 \right).\label{Hamiltonian}
\end{align}

The photon contribution to the magnetization  has the form

\begin{equation}
{\mathbf{\mathcal M}}_w=
%\mathcal{L}_{\mathcal{F}}{\bf B}_w
\mathcal{L}_{\mathcal{GG}} ( {\bf E}_w \cdot {\bf B}_e) {\bf E}_{w,3}  +\mathcal{L}_{\mathcal{FF}} ({\bf B}_w \cdot {\bf B}_e ) {\bf B}_{w,3}.
%({\bf B}_e +{\bf B}_w).
\label{magnetizationf1}
\end{equation}

The magnetization depends on the polarization mode and provided values of the Lagrangian derivatives and it is always positive.This induces a paramagnetic nature for the photon.
%More specifically, for modes (2) and (3) adopts the form
%\begin{align}
%{\mathcal E}_{mag}^{(2)}=&-
%{\mathcal L}_{\mathcal F} ({\bf B}_w\cdot
%{\bf B}_e )+
%\mathcal{L}_{\mathcal{GG}}({\bf B}_e\cdot{\bf E}_w)^2,  \\
%+ \mathcal{L}_{\mathcal{GG}}(\bf{B}_e\cdot\bf{E}_w){\b%f E}_{w,1}{\bf B}_{w,1}, \\
%{\mathcal E}_{mag}^{(3)}=&-
%\mathcal{L}_{\mathcal{FF}} ({\bf B}_e\cdot{\bf B}_w)^2.
%{\bf B}_eB_{w,3}.
%\end{align}

Considering photon oscillating fields $B_w= E_w= E_0 e^{i(\omega t-k_{\bot} y)}$  in Eq.(\ref{magnetizationf1}) %with ${\bf E}_0={\bf E}_{0,\parallel}\hat{{\bf x}}_{\parallel}+{\bf E}_{0,\bot}\hat{{\bf x}}_{\bot}$,
and from space-time averages $\langle \mathcal{M}_{w} \rangle$, only even powers of wave fields will give a non-vanishing contribution.  Further, when considering effective measurable energy values we must actually integrate over volume, i.e. compute $\langle \mathcal{E}_{mag} \rangle_V$.

This allows us to compute  the effective magnetic moment of a photon probe  defined as
\begin{equation}
\mid \pmb{\mu}_{ph}\mid=-\frac{d \langle \mathcal{E}_{mag} \rangle_V } {d B_e} \frac{1}{V\langle N_V\rangle},
\end{equation}
where $N_V^{(i)}$ is the number density of ith mode  ($i=2,3$) and is given by  $N_V^{(2,3)}= \frac{1}{2}\frac{E_0^2}{\omega^{(2,3)}}$ \cite{Villalba-Chavez:2012pmx}. $\pmb{\mu}_{ph}$ will characterize the interaction of propagating photons with the vacuum virtual pairs under the presence of a  magnetic field. More specifically for the two modes $\mid \pmb{\mu}_{ph}^{(2,3)} \mid$ reads as

\begin{align}
&\mid\pmb{\mu}_{ph}^{(2)}\mid =\dfrac{\alpha}{16\pi}\dfrac{1}{b^3} \left\lbrace
3-12 \zeta^{(1,1)}\left( -1,\dfrac{1}{2b} \right )+3\psi\left( \dfrac{1}{2b} \right )\right\rbrace.\nonumber\\
&+ b \left[-3 +\log  \Gamma \left(\dfrac{1}{2b}  \right ) \left (\dfrac{\pi}{b}
\right )^2\right. +\left.\left.\psi^{(1)}\left (1+\dfrac{1}{2b}\right )+ 2b^2\right ] \right\rbrace\dfrac{\mid\bf{k}_{\bot}\mid}{B_c},
\end{align}

\begin{align}
&\mid \pmb{\mu}_{ph}^{(3)}\mid =\dfrac{\alpha}{8\pi}\dfrac{1}{b^4}\left\lbrace
	-\psi^{(1)}\left (1+\dfrac{1}{2b}\right ) \right.
	+ b \left [	4-	4\psi\left (1+\dfrac{1}{2b}\right )+2\psi\left (\dfrac{1}{2b}\right )
	\right  ] \nonumber\\
&+b^2 \left[4-2 \log(2\pi)
+\left.\left. 4\log\left (\Gamma \left( \dfrac{1}{2b}  \right ) \left (\dfrac{\pi}{b}  \right )^{1/2}\right ) \right] \right\rbrace \dfrac{\mid{\bf{k}_{\bot}}\mid}{B_c},  \right.
\end{align}

\noindent where now $b=B_e/B_c$,  $\psi^{(1)}=\partial_h\psi[h]$ being $\psi$  the PolyGamma o Digamma function, (first derivative of ${\rm ln}\, \Gamma$). $\zeta^{(1,1)}[s,h]=\partial_h \zeta^{\prime}$ with $\zeta^{\prime}=\partial_s\zeta[s,h]$ and $\zeta[s,h]$ is the Hurwitz zeta function  \cite{Adam}.
In appendix (\ref{apendice:1}) we provide expressions for $\psi$ and  $\zeta^{\prime}$ in weak and strong field limits.

The photon effective magnetic moment inherits the orientation of the magnetization. For both modes, it is parallel to the external magnetic field,  determining the photon field component. In the  weak field limit, the photon effective magnetic moment yields for modes (2) and (3) %(see Appendice (\ref{apendice:1}),
\begin{align}
\mid \pmb{\mu}_{ph}^{WF\,(2)}\mid &=\frac{7}{2} \xi\mid  \pmb{k}_{\bot}\mid B_e =\frac{\alpha}{4\pi}\frac{28}{45} \frac{B_e}{B_c^2}\mid \bf{k}_{\bot}\mid,\\
\mid \pmb{\mu}_{ph}^{WF\,(3)}\mid &=2 \xi\mid \pmb{k}_{\bot}\mid B_e =\frac{\alpha}{4\pi}\frac{8}{45}\frac{B_e}{B_c^2}\mid{\bf k}_{\bot}\mid.
%\frac{\alpha^2}{4\pi}\frac{28}{45}\frac{k_{\bot}}{m^2}
\end{align}

Instead, in the strong field limit only  mode (2) contributes to it tending  to a constant value found to be a thousand times smaller than the anomalous electron magnetic moment, $\mu_e$,
\begin{align}
\mid \pmb{\mu}_{ph}^{SF\,\,(2)}\mid=\frac{\alpha}{3\pi}\frac{\langle E_w^2 \rangle}{B_c}&\sim \frac{\alpha }{3\pi}\frac{e}{2m_e}\frac{\mid {\bf k}_{\bot}\mid}{m_e}\sim 10^{-3}\mu_e, \nonumber\\
\mid \pmb{\mu}_{ph}^{SF\,\,(3)} \mid&=0.
\end{align}
\noindent

As was pointed out in  \cite{Villalba-Chavez:2012pmx} performing an analogy between the parallel magnetic moment for a photon probe  $\pmb{\mu}_{ph}$ and the spin magnetic moment of the electron, the former could be interpreted  as a consequence of the existence of spin for a photon probe. However, it is merely an analogy, since spin for particles has to obey additional quantum mechanical properties that the bosonic photon field does not satisfy.
%%%%%%%%%%%%%%%%%%%%%%%%%%%%%%%%%%%%%%%%%%%%%%%%
\begin{figure}[t]
\centering
  \includegraphics[width=0.8\linewidth]{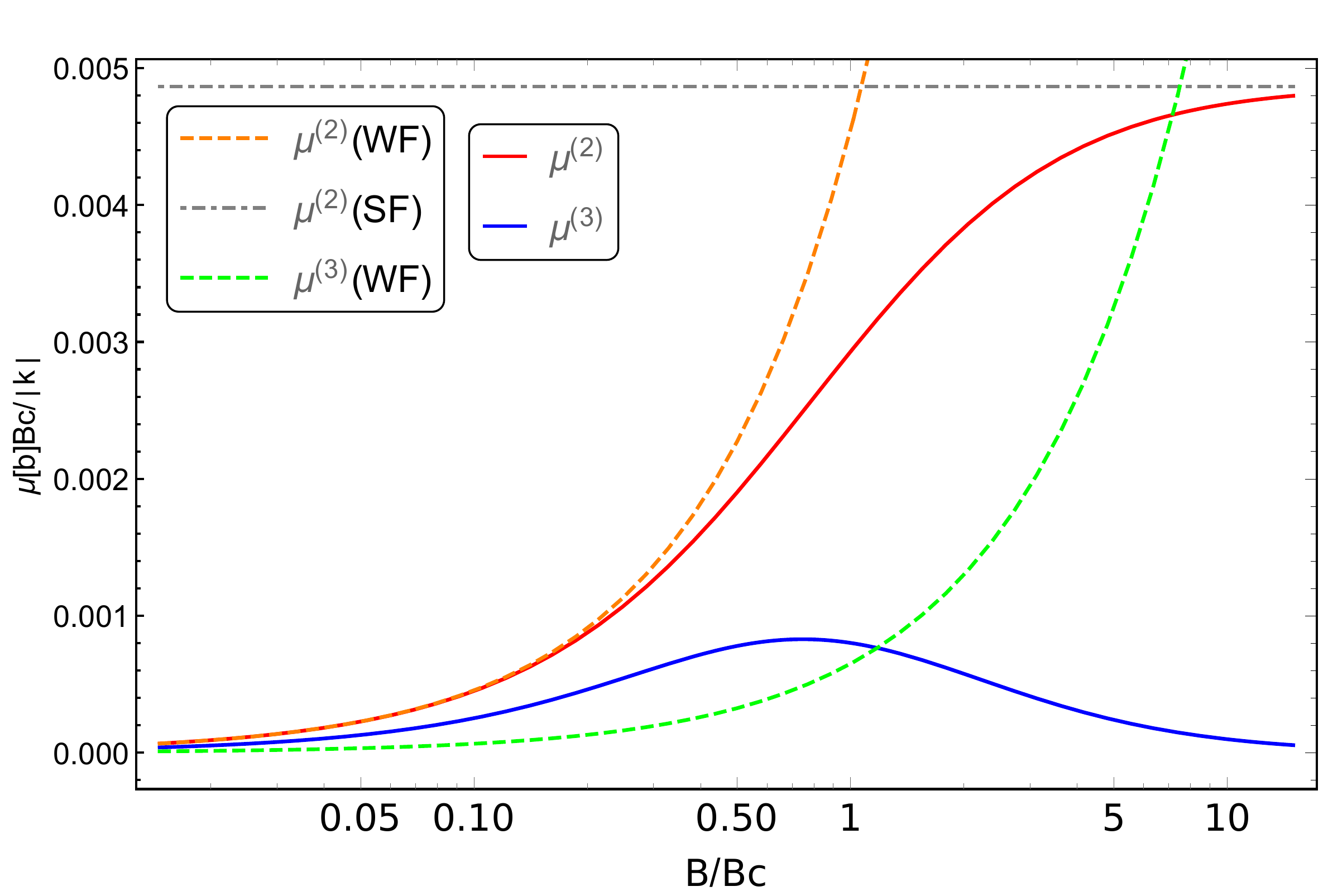}
 \caption{Photon effective magnetic moment as a function of magnetic field strength, $b=B/B_c$, for mode (2) (mode (3)) is shown in red (blue) solid line. Weak (strong) field limit is depicted in  orange dashed (grey  dot-dashed) line for mode (2). The weak field limit for mode (3) is depicted by the dashed green line. }
\label{fig:MM}
\end{figure}
%%%%%%%%%%%%%%%%%%%%%%%%%%%%%%%%%%%%%%%%%%%%%
In Fig. (\ref{fig:MM}) we plot the photon effective  magnetic moment as a function of dimensionless parameter $b=B/B_c$ for light polarization modes (2) and (3).
For comparison, we have depicted the magnetic moment in the weak and strong field limits as well. One can see that the effective magnetic moment for mode (2) (model (3)) is over (under) estimated if we take the approximate values arising from the weak and strong limits in their range of validity. Then, for a magnetic field $B_e \gtrsim  2B_c$ the effective magnetic moment of photons polarized on mode (2) tends asymptotically to a constant value. Instead, photon magnetic moment for mode (3) slowly decreases  with the magnetic field strength, being zero its strong field limit value.  Let us note that the effective magnetic moment for mode (2) coincides with those obtained  by \cite{Mielniczuk} for arbitrary values of the magnetic field strength and \cite{Villalba-Chavez:2012pmx} in the strong field limit.

As a complementary remark, let us mention that the magnetic moment of the  electron in presence of a magnetic field considering radiative corrections in QED, was studied in \cite{Ferrer:2015wca} for weak and strong magnetic field limits providing expressions $\sim \frac{\alpha}{2\pi}\frac{e B}{2m_e}$ and  $\sim \frac{m_e \alpha}{4\pi}{\rm ln^2} (\frac{m^2_e}{2 e B})$, respectively. So far, there is no general expression for the whole range of magnetic field strengths. However, our study for photon effective magnetic moment for arbitrary values of the magnetic field  contributes to clarifying the physical implications of those limits.

The physical picture that emerges for the effective magnetic moment of photons is thus a consequence of their propagation in a magnetized vacuum. Likewise photons in a medium, photons in vacuum behave as quasi-particles interacting with virtual pairs and external magnetic fields via this effective moment.

\section{Experimental prospects for testing vacuum properties with photon propagation with EH NLED}
\label{sec:exp}

Indirect  observation of photon magnetization and magnetic moment are possible in the astrophysical scenario, as already shown in the literature \cite{Mielniczuk,Turolla:2017tqt}.
%In a different scenario, astrophysical observations could be used to estimate the effective magnetic moment.
When photons cross the magnetic field in the magnetosphere of a NS they are deflected and their trajectories are bent, rising magnetic lensing effects where a non-vanishing photon magnetic moment plays an important role, see  \cite{PhysRevLett.94.161101}. Instead, on Earth, much lower strength pulsating fields can be generated in modern laser facilities on tabletop experiments \cite{PVLAS:2007wzd,Tommasini:2009nh} constituting yet another interesting opportunity to measure quantities such as photon magnetization, photon effective magnetic moment or photon pressure in the weak field limit, as discussed.
As we have shown, photon magnetization and effective magnetic moment are mainly determined by
the  magneto-electric terms in the Lagrangian $1/2\mathcal{L}_{\mathcal{GG}}({\bf B}_e \cdot {\bf E}_w)^2$ and  $1/2\mathcal{L}_{\mathcal{FF}}({\bf B}_e\cdot {\bf B}_w)^2$ depending on the polarization mode. In the EH weak field limit  ${\mathcal L}_{\mathcal {GG}} B_e^2 \rightarrow 7/4 \xi B_e^2$ while   ${\mathcal L}_{\mathcal {FF}} B_e^2\rightarrow 2\xi B_e^2$, respectively.
Therefore, experiments aimed to get  magnetization properties, even if challenging, would give us the $\mathcal{L}_{\mathcal{GG}}$ and  $\mathcal{L}_{\mathcal{FF}}$ values independently.

As already mentioned, the photon pressure  ($\pmb{{\hat y}}$-direction component)
$p_w^2$ for modes (2,3) is  written as  $p_w^2=P_0+C^{2,3}\mathcal{M}_w^{(2),(3)}B_e$ with 
$P_0=(1-\mathcal{L}_{\mathcal{F}})\left\langle E_w^2\right\rangle$ increased in a quantity proportional to the magnetization of the photon ($C^{2,3}=1/2,-3/2$) i.e. for each polarization mode will be different and likely better suited to discriminate over the Cotton Mouton birefringence usually quoted. The latter is sensitive to the refraction index difference $\Delta n$ related to the difference $(\mathcal{L}_{\mathcal{GG}}-\mathcal{L}_{\mathcal{FF}})B_e^2$.
The classical pressure diminishes (increases) depending on the mode in a quantity proportional to the magnetic energy ($\sim \mathcal{M}_w.B_e$).
Therefore, in principle, the radiation pressure could be accessible to experimental setups. An ideal perfect mirror-plate along the plane $xz$ (parallel to the magnetic field) may feel the radiation pressure  produced by a laser beam, with $\langle E_{w}^2\rangle\sim I$. This pressure for mode 2 (mode 3) will be higher (lower) than the  classical (corresponding to photons travelling in an ``empty vacuum'') \cite{radpres,Mahrle}.

Interestingly, experimental setups based on light scattering \cite{Ferrando:2007pgk} and Dirac materials \cite{diracmaterials} have already obtained measurements of magnetization that seem promising. For the latter their associated values of Schwinger fields are $E_c\sim 10^5$ V/cm and $B_c\sim 1$ T, both being  experimentally accessible and providing a platform to explore the strong field regime of QED and to observe a new class of magneto-electric effects such as a high electric field modulated magnetization and a very large enhancement of the dielectric constant.  These effects are also highly anisotropic as they depend on relative orientation of E/B fields and their crystallographic orientation

Considering photon probes from a laser beam whose intensity is related with electric and magnetic field of the photon by $\frac{I}{c}=\epsilon_0 E_w^2=\frac{B_w^2}{\mu_{0}}$,  typical current values of magnetic field generated in laboratory attain around $\sim 45 \,\rm T=4.5\times 10^5$ G, while future intensities of $I=10^{28}$ $\rm W/m^2$, in CGS units  $I=10^{24}$ $\rm W/cm^2$, will require $\sim 10^2$ PW lasers.  Very recent advances on this side \cite{megaT} include the possibility of generating ultrahigh magnetic fields of the MegaTesla order using microtube implosions driven by ultraintense and ultrashort laser pulses in  a novel scheme.
 For a future benchmark intensity, $I=10^{24}$ $\rm W/cm^2$ and $B_e \sim 30 \,\rm T=3\times 10^5$ G, we can estimate the theoretical vacuum magnetization for both photon modes
$\mathcal{M}_w^{(2)}=\frac{7}{2} \frac{\xi I}{c} B_e \sim 2.3 \times 10^{-4}\,\rm  G$ and ${\mathcal M}_w^{(3)}=2  \frac{\xi I}{c} B_e  \sim 1.3 \times 10^{-4}\,\rm G$.

To illustrate with numbers, at atmospheric conditions of pressure $P$ and temperature $T$, one can obtain the estimate of Inverse Cotton Mouton (ICM)  magnetization,  $\mathrm{M}_{ICM}^{\text {atom}}$ finding these values  are two  orders of magnitude higher than those
for Helium magnetization \cite{Rizzo:2010di} at reach in current facilities \cite{Karbstein:2019oej,King}. However, we should keep in mind that ionization appears at different intensities, typically $I>10^{19}$ $\rm W/m^2$ depending on the noble gas and the consequent ion current could somewhat perturb the ICM measurement. Recently, on this same line, it has been  stated \cite{Roso} that performing photon-photon collision experiments  using two counterpropagating laser beams could yield very promising results for intensities close to  $I\sim 10^{24}$ $\rm W/cm^2$.

\section{Conclusions}
\label{conclusions}
Starting from the EH NLED, the EMT and AMT associated to a photon probe, propagating transverse to a constant and uniform background magnetic field have been calculated in vacuum, using the  Einstein-Hilbert method. Our calculations have been performed using NLED in  the soft-photon regime which assures the validity of our results up to $B_e \sim 430 B_c$. Thus for applications in astrophysical scenarios or studies related to ultraintense laser experiments, concerning strengths below the critical field, $B_c$, our findings remain valid. In particular energy density, pressures, Poynting vector, angular momentum, magnetization, and photon effective magnetic moment have been derived for those strengths in a more general and accurate fashion for the selected NLED, rather than using the usual weak and strong field limits.

By  construction the robust and elegant  Euler-Hilbert  method  guarantees symmetric, conserved, and gauge-invariant EMT. Besides, the EMT found, is anisotropic and lacks tracelessness.
Besides, we have confirmed that, in general, the assumption of the equivalence of EMT from the Hilbert method  and that by improved Noether technique used in \cite{Villalba-Chavez:2012pmx} is not fulfilled.
We have also identified this effect originating in the non-linear magneto-electric terms of the Lagrangian (second order in the photon fields) and the convenience of the Hilbert method to obtain a symmetric gauge invariant EMT. 
Nevertheless, despite finding that physical magnitudes in Noether and Hilbert procedures thus differ, some general properties of the non-linear theory like EMT anisotropy and not being traceless are valid for both formalisms. As EMT is anisotropic, the rotational symmetry is broken and is replaced by an axial symmetry driven by the direction of the external magnetic field. 
We compare the Hilbert and Noether methods and discuss some observables. 
The non symmetric EMT obtained `` \`a la Noether'' yields non conservation of angular momentum vector and the appearance of a finite torque and perpendicular magnetization. Besides,  pressures obtained using the Hilbert or Noether method are different for each polarization mode. In addition,  we find the photon radiation does not fulfill that $p_w^2=\mathcal{P}_w$, as occurs for photon propagating in isotropic media or in ``empty vacuum''. 

Therefore, the photon probe feels the  non-linear magnetized vacuum effects. In this qualitative sense, vacuum anisotropic pressures have some analogy with the Casimir effect. Both effects lead to differences in pressures and bear a common origin, axial symmetry. The former is determined by the external magnetic field and the latter due to boundary conditions: the presence of plates in Casimir effect \cite{faccio,Elizabeth,Marklund:2008gj}.  

Let us remark that the results obtained by the Noether and Hilbert methods can reconcile for the EH NLED i.e,  giving the same results,  if the Noether EMT is calculated not starting off from  ${\mathcal L}^{\rm (ph-B)}$ but from a general Lagrangian $\mathcal{ L}(\mathcal {F},\mathcal {G})$  and then performing the expansion up to second order in the photon fields, on the resulting symmetric EMT.  

In addition, we have also studied the magneto-electric properties of the magnetized vacuum finding that it behaves as  paramagnetic. A photon effective magnetic moment is also calculated. It agrees for mode (2) with the result obtained by \cite{Mielniczuk}. In the weak field limit, we reproduce the  linear behavior of the magnetic field obtained in complementary way from radiative corrections of QED and the corresponding  dispersion equations \cite{Elizabeth}. Instead, for strengths, $B_e\gg 2B_c$, the effective photon  momentum tends to another constant value %$\frac{e}{3\pi}(\frac{e}{2m})$
in accordance to results obtained in the strong magnetic field limit \cite{Villalba-Chavez:2012pmx,Mielniczuk}. Finally, based on previous works regarding experiments proposed by \cite{Ferrando:2007pgk,Tommasini:2009nh} and promising measurement of vacuum magnetization of Dirac materials we have discussed the possibility of measuring magnetization of vacuum, photon radiation pressure in future experiments although they present important challenges.

\section*{Acknowledgments}
The authors thank to C. Albertus, L. Roso, L. Volpe, A. W. Romero Jorge for their  invaluable remarks and discussions. A.P.G. and A.P.M. acknowledge the support of the Agencia Estatal de Investigaci\' on through the grant PID2019-107778GB-100 and from Junta de Castilla y Le\'on, SA096P20 project. The work of E.R.Q. was  supported by the project of No. NA211LH500-002 from AENTA-CITMA, Cuba. This manuscript has no associated data
or the data will not be deposited.

\newpage
\section{Appendix}
\label{apendice}

\subsection{Derivatives of the effective EH Lagrangian}\label{apendice:1}

As mentioned, in section \ref{section:1} the Lagrangian derivatives can be calculated for arbitrary external magnetic field $B\equiv B_e$ and the result reads as follows \cite{DittrichLibro}

\begin{align}\label{LFLFFLGG}
&\mathcal{L}_{\mathcal{F}}=-\mu_{0}^{-1}-\frac{\alpha}{2 \pi \mu_{0}}\int_{0}^{\infty} d z e^{-\frac{B_{\text {c }}}{B} z}\nonumber\\
&\times\left[-\frac{2}{3 z}-\frac{1}{z \sinh ^{2}(z)}+\frac{\operatorname{coth}(z)}{z^{2}}\right],\\
\mathcal{L}_{\mathcal{FF}}&=-\frac{\alpha}{2 \pi \mu_{0}B^2} \int_{0}^{\infty} d z e^{-\frac{B_{\text {c }}}{B} z}\nonumber\\
&\times \left[2\frac{z\operatorname{coth}(z)-1}{z \sinh ^{2}(z)}
+\frac{1}{z \sinh ^{2}(z)} -\frac{\operatorname{coth(z)}}{z^{2}}\right],\\
\mathcal{L}_{\mathcal{GG}}&=-\frac{\alpha}{2 \pi \mu_{0}B^2}\int_{0}^{\infty}dze^{-\frac{B_{\text {c}}}{B} z}\nonumber\\
&\times \left[-\frac{2}{3}\operatorname{coth} (z)
-\frac{1}{z \sinh ^{2}(z)}+\frac{\operatorname{coth}(z)}{z^{2}}\right].
\end{align}

In the limit $E\rightarrow 0$, i.e. $b\rightarrow0$ we can solve these integrals using functional regularization so that using $\mathcal{L}^{EH}$ in Eq. (\ref{eqEH})
%for Eqs. (9)-(11)

\begin{align}
&\mathcal{L}_{\mathcal{F}}=\frac{\partial\mathcal{L}^{EH}}{\partial\mathcal{F}}\mid_{f=0}=-\mu_{0}^{-1}-
\frac{\alpha}{2 \pi\mu_{0}}\left(\frac{1}{3}+2 h^{2}-8 \zeta^{\prime}(-1, h)\right.\nonumber \\
&+4 h {\rm ln} \Gamma(h)-2 h {\rm ln} h+\left.\frac{2}{3} {\rm ln} h-2 h {\rm ln} 2 \pi\right),\nonumber \\
&\mathcal{L}_{\mathcal{FF}}=\frac{\partial^2\mathcal{L}^{EH}}{\partial^2\mathcal{F}}\mid_{f=0}\nonumber \\
&=\frac{\alpha}{2\pi \mu_{0}B^2} \left(\frac{2}{3} +4 h^2 \psi(1+h)-2h-4h^2-4h {\rm ln} \Gamma(h)\right. \nonumber \\
&+2h {\rm ln} 2\pi -2h{\rm ln}h\left.\right), \nonumber \\
&\mathcal{L}_{\mathcal{GG}}=\frac{\partial^2 \mathcal{L}^{EH}}{\partial^2 \mathcal{G}}\mid_{f=0}\nonumber \\
&=\frac{\alpha}{2\pi\mu_{0}B^2}\left(-\frac{1}{3}-\frac{2}{3}\left(\psi(1+h) -2h^2 +(3h)^{-1}\right)\right. \nonumber \\
 &+\left.8\zeta^{\prime}(-1,h)-4h {\rm ln} \Gamma(h) + 2h {\rm ln} 2\pi +2h {\rm ln} h\right), \label{derivativesB}
\end{align}

%\begin{align}
%(1-L_F)=\left (1-\frac{\alpha}{2\pi \mu_{0}}\left (  %\frac{1}{3} +2h^2-8 {\zeta}^{\prime} (-1,h) %+4h\Gamma(h)-2hln h +\frac{2}{3} ln h-2h ln2\pi\right %)\right ),\\
%L_{FF}=\frac{\alpha}{2\pi \mu_{0}B^2}\left (\frac{2}{3} + 4 %h^2 \psi(1+h) -2h-4h^2-4h ln \Gamma(h) +2h ln (2\pi) -2hln %(h)\right ), \\
%L_{GG}=\frac{\alpha}{2\pi \mu_{0}B^2}\left (\frac{1}{3}-\frac{2}{3}\left ( \psi(1+h) +2h^2 +(3h)^{-1}\right ) +8 {\zeta}^{\prime} (-1,h)-4hln \Gamma(h)\right ),\\
\noindent where  $h=\frac{B_c}{2B_e}$, and the quantum corrections are  proportional to the fine structure constant,  $\alpha$. $\psi$ denotes the PolyGamma o Digamma function (first derivative of ${\rm ln}\, \Gamma$) and $\zeta^{\prime}$ is the first derivative of the Hurwitz zeta function with respect to the first argument.

{ Let us remark that the expansion on the derivatives of the effective Lagrangian $\mathcal{L}^{EH}$ and their integration Eq. (\ref{derivativesB}), for $\mathcal{G}=0$ could be  extended for the study of the propagation of photon:  in a pure background  electric field and/or
in the background of an orthogonal electric and magnetic field, with  $h\rightarrow \frac{B_c}{2\sqrt{2\mathcal{F}}}$. }

For the weak field  (WF) case $h>1$ the functions ${\zeta}^{\prime} (-1,h)$, ${\rm ln}\,\Gamma(h)$ and $\psi(1+h)$ have the asymptotic  expressions

\begin{align}
 &{\zeta}^{\prime} (-1,h)=\frac{1}{12}-\frac{h^2}{4}+\frac{{\rm ln} h}{2}B_2(h) \nonumber\\ &+\underbrace{\int_{0}^{\infty}\frac{e^{-hx}}{x^2}\left ( \frac{1}{1-e^{-x}}-\frac{1}{x}-\frac{1}{2}-\frac{x}{12}\right )}_{\frac{1}{720}\frac{1}{h^2}}, \quad Re (h) >0\nonumber\\
&{\rm ln}\Gamma(h)=-\frac{{\rm ln} h}{2}+\frac{1}{12 h}+\frac{1}{360 h^2} +h{\rm ln} h -h-\frac{1}{2}+\frac{1}{2}{\rm ln} 2\pi,\nonumber\\
&\psi(1+h)={\rm ln} h +\frac{1}{2h}-\frac{1}{12h^2}+\frac{1}{120h^4},
\end{align}
and $B_2(h)=h^2-h+\frac{1}{6}$ is the second Bernoulli polynomial  %\cite{StirlingMod}.
Then, $\mathcal{L}_{\mathcal{F}}$, $\mathcal{L}_{\mathcal{FF}}$
and $\mathcal{L}_{\mathcal{GG}}$ become

\begin{align}
&\mathcal{L}_{\mathcal{F}}=-1-\frac{\alpha}{2\pi}\left(\frac{1}{3} +2h^2-8\left(\frac{1}{12}
-\frac{h^2}{2}+\frac{{\rm ln}h}{2}B_2(h)+\frac{1}{720h^2}\right)\right.\nonumber  \\
&+4h\Gamma[h]+2h{\rm ln}(2\pi) + 2h{\rm ln}h \left.\right),\label{LFLFFLGGex}\\
&\mathcal {L}_{\mathcal{GG}}=\frac{\alpha}{2\pi B_e^2}
\left(\right.-\frac{1}{3}-\frac{2}{3}\left (\psi(1+h) +2h^2 +\frac{1}{3h}\right )\nonumber\\
&-2h\left(\frac{1}{6h}-\frac{1}{180h^3}\right )-\frac{2}{3}\rm{ln}(h)- 4h
\left (\frac{{\rm ln}(h)}{6}+\frac{1}{360h^2}\right )\left.\right),\nonumber\\
&\mathcal {L}_{\mathcal{GG}}\xrightarrow{WF}\frac{\alpha}{2\pi B_e^2}\left (\frac{1}{18h^2}+\frac{1}{90h^2}+\frac{1}{360h^2 }\right )=\frac{7}{2}\frac{8}{45}\frac{\alpha}{4\pi B_c^2}=\frac{7}{2}\xi,\\
&\mathcal{L}_{\mathcal{FF}}=\frac{\alpha}{2\pi B_e^2}\left(\frac{2}{3}-2h\left(\frac{1}{6h}-\frac{1}{180h^3}\right)+6h^2\left ( \frac{2}{3}\psi(1+h) +2h^2 +\frac{1}{3h}\right )\right )\nonumber\\
&\mathcal {L}_{\mathcal{FF}}\xrightarrow{WF}\frac{\alpha}{2\pi B_e^2}\left (\frac{1}{90h^2}+\frac{1}{90h^2}
\right )=\frac{\alpha}{2\pi B_e^2}\left (\frac{4}{90h^2}\right )=\frac{8}{45}\frac{\alpha}{4\pi B_c^2}=2\xi,\label{LF}
\end{align}

The asymptotic behavior, i.e. $h<1$ i.e. strong field (SF) limit of functions  ${\zeta}^{\prime} (-1,h)$, ${\rm ln}\,\Gamma(h)$ and  $\psi(1+h)$ reads

 \begin{align}
 {\zeta}^{\prime} (-1,h)&=-\frac{h^2}{4}+\frac{{\rm ln} h}{2}B_2(h),\\
{\rm ln}\,\Gamma(h)&=-\frac{{\rm ln} h}{2}+h{\rm ln} h -h+\frac{1}{2}{\rm ln} 2\pi,\\
\psi(1+h)&=\gamma +\frac{\pi^2}{6}h-\zeta(3)h^2...,
 \end{align}
\noindent and in this limit  $\mathcal{L}_{\mathcal{F}}$, $\mathcal{L}_{\mathcal{FF}}$ and $\mathcal{L}_{\mathcal{GG}}$ have the form
\begin{align}
\mathcal{L}_{\mathcal{F}}=\frac{\alpha}{3\pi} {\rm ln}\, (B/B_c),\\
\mathcal{L}_{\mathcal{GG}}=\frac{\alpha}{3\pi B_e B_c},\\
\mathcal{L}_{\mathcal{FF}}=\frac{1}{B_e^2}\frac{\alpha}{3\pi}.
\end{align}

\subsection{Maxwell equations for magnetized vacuum in EH NLED}
\label{apendice:0}
We summarize the equations of motion and dispersion laws for a photon probe in the presence of an external fixed background magnetic field. It can be extracted directly from the Lagrangian in Eq. (\ref{lagfF}) presented before. We first select the direction of a photon probe propagating in $\pmb{\hat{y}}-$direction,  transverse to a fixed constant magnetic field along the $\pmb{\hat{z}}-$direction, ${\bf B}=B(0,0,1)$. Therefore the equation of motion is determined using Euler-Lagrange equations $\frac{1}{\sqrt{-g}}\partial_{\nu} \left[ \sqrt{g} { F}^{\mu \nu}\right]=0$.

We can express the non-linear Maxwell equations in the local flat geometry in the more familiar form focusing on the wave fields using the electric displacement field. $\mathbf{D}_w$, and magnetic field, $\mathbf{H}_w$,
\begin{align}
\frac{\partial\mathbf{D}_w}{\partial t}=-\nabla \times \mathbf{H}_w,   \quad \nabla \cdot \mathbf{D}_w=0, \label{magpo1}
\end{align}
\noindent and a second pair for ${\bf E}_w$ and ${\bf B}_w$
\begin{align}
\nabla \cdot \mathbf{B}_w=0,\quad
\frac{\partial \bf{B}_w}{\partial t}=-\nabla \times  \mathbf{\bf{E}}_w, \label{magpo2}
\end{align}
\noindent with the constitutive equations for $\mathbf{D}_w$ and $\mathbf{H}_w$ components being
\begin{align}
D_{w,i}&=\frac{\partial}{\partial E_{w,i}}[ \mathcal{L}^{\rm (ph-B)}]=\epsilon_{ij}E_{w,j}=E_{w,i} + P_{w,i},\label{dw} \\
H_{w,i}&=-\frac{\partial}{\partial B_{w,i}}[ \mathcal{L}^{\rm (ph-B)}]=(\mu^{-1})_{ij}{B}_{w,j} =B_{w,i}-M_{w,i}, \label{mag}
\end{align}
with $i=1,2,3$. Similar to an optical medium, ${\bf P}_w$ and ${\bf M}_w$ are the resulting  polarization and magnetization of the photon probe due to the magnetized vacuum. Note that the presence of an arbitrary strength external magnetic field will impact not only the electric permittivity and magnetic permeability tensors, $\epsilon (B)$ and $\mu^{-1}(B)$, but also Maxwell equation solutions for the photon fields in ${\bf D}_w$ and ${\bf H}_w$ thus affecting its propagation in a magnetized vacuum.
%%%%%%%%%%%%%%%%%%%%%%%%%%%
%\begin{figure}[h!]
%\centering
%\includegraphics[width=0.95\linewidth]{Modes2y3.pdf}
% \caption{Polarization modes for a photon. We depict the case  with parallel wave and external magnetic fields, $\bf{B}_w \parallel B_e$, i.e.mode 3 (right panel) and those perpendicular $\bf{B}_w \bot \bf{B}_e$ i.e. mode 2 (left panel).}
%  \label{fig:modes}
%\end{figure}
%%%%%%%%%%%%%%%%%%%%%%%%%%%%

In our configuration, explicit expressions for the $\epsilon$ and $\mu^{-1}$  components can be expressed in terms of Lagrangian derivatives as
\begin{align}
\epsilon_{11}=(\mu^{-1})_{11}=1-\mathcal {L}_{\mathcal F},\label{epsilon}\\
\epsilon_{22}=(\mu^{-1})_{22}=1-\mathcal {L}_{\mathcal F},\\
\epsilon_{33}=(1-\mathcal {L}_{\mathcal F}+2{\mathcal F}\mathcal {L}_{\mathcal {GG}}),\\
(\mu^{-1})_{33}=(1-\mathcal {L}_{\mathcal F}-2{\mathcal F}\mathcal {L}_{\mathcal {FF}}),\label{epsimu2}
\end{align}
\noindent being zero otherwise. Further, we define in our system $\epsilon_{\parallel}=\epsilon_{33}$, $\epsilon_{11}=\epsilon_{22}=\epsilon_{\bot}$, $\mu_{33}=\mu_{\parallel}$, $\mu_{11}=\mu_{22}=\mu_{\bot}$.
From Maxwell equations in Eqs.(\ref{magpo1}),(\ref{magpo2}) for ${\bf D}_w$ and ${\bf H}_w$ and assuming a plane wave photon field in the form ${\bf E}_w={\bf E}_0 \rm exp[-i(k_{\perp}y-\omega t)]$, propagating in the $\pmb{\hat{y}}-$direction we obtain the dispersion equation in Fourier space for a monochromatic wave as
\begin{equation}
(\epsilon_{ijk}\epsilon_{lab}k_{j}(\mu^{-1})_{kl}k_a+\omega^{2}\epsilon_{ib})E_{w\,b}=0,\label{ecdis}
\end{equation}

\noindent where $i,j,k,a,b,l=1,2,3$. Solutions to the previous Eq.(\ref{ecdis}) describe two physical polarization modes of the photon field, (2) and (3).
%In Fig.(\ref{fig:modes}) we show the arrangement:
{ Mode (2) with ${\bf B}_w \bot {\bf B}_e$ ${\bf E}_w \parallel {\bf B}_e$ and mode (3) with ${\bf B}_w \parallel {\bf B}_e$, ${\bf E}_w \bot {\bf B}_e$.}
If we now set the wave number $k_{\bot}=k_2$ assuming propagation in the $\pmb{\hat{y}}$-direction we obtain linear birefringence  due to the linear polarization of the radiation

\begin{align}
\omega^{(2)}&\simeq c\mid {\bf k}_{\bot} \mid (1-\frac{{\mathcal L}_{\mathcal {GG}}\bf{B}_e^2}{2}),\\
\omega^{(3)}&\simeq c\mid{\bf k}_{\bot} \mid (1-\frac{{\mathcal L}_{\mathcal {FF}}\bf{B}_e^2}{2}).
\end{align}
\noindent in line with \cite{Elizabeth,Shabad:2011hf}.
 Besides,  Cotton Mouton birefringence \cite{Rizzo:2010di} appears and the  refraction index is associated to the two different polarizations modes: $n_{\parallel}$ for mode (2)  and $n_{\bot}$ for mode (3)
\begin{equation}\label{indrefraccion}
n_{\parallel,\bot}=\frac{\mid \bf{k_{\bot}}\mid }{\omega^{(2,3)}}=\sqrt{\frac{\epsilon_{\parallel,\bot}}{\mu_{\bot,\parallel}}}.
\end{equation}

\noindent If we now define its difference as
\begin{equation}
\Delta n=\frac{(\mathcal{L}_{\mathcal{GG}}-\mathcal{L}_{\mathcal{FF}})\bf{B}_e^2}{2}.
\end{equation}
 In the weak field limit (see appendix \ref{apendice:1}) it is reduced to the well-known result $\Delta n^{\rm WF}_{CM}=3/4\xi B_e^2$, instead for the strong limit $\Delta n^{\rm SF}_{\rm CM}=\frac{\alpha}{6\pi}(\frac{B_e}{B_c}-1)$. Since the velocity
 $v_{\parallel,\bot}=1/n_{\parallel,\bot}$, in the strong field limit the condition, $v_{\parallel}^{SF}=1-\frac{\alpha}{6\pi}(\frac{B_e}{B_c})>0$ ($\alpha/(4\pi)=1/137$)  fixes the validity of one-loop approximation up to values of the magnetic field  $B_e/B_c\leq \pi/\alpha\sim 430$  \cite{DittrichLibro}.
 %\apgcomm{REVISAR, no es consistente con  ec 21 en SF}

If we particularize the previous for electric permittivity and magnetic permeability in weak and strong field limits we obtain the following expressions.
%\label{apendice:2}
%NO SE USA

%Once,  was defined the theory, we summarize the main properties of the %propagation of photon in a vacuum
In the weak field limit magnetic permeability components  in presence of an external magnetic field $\textbf{B}_e=B_e \pmb{\hat{z}}$ are

\begin{align}\label{epsilon2}
\epsilon_{11}=(1-\xi B_e^2),\\
\epsilon_{33}=(1+\frac{5}{2}\xi B_e^2),\\
(\mu^{-1})_{11}=(1-\xi B_e^2),\\
(\mu^{-1})_{33}=(1-3\xi B_e^2),
\end{align}
and refraction indices and quadratic velocities in parallel and perpendicular components as
\begin{align}
n_{\parallel}=1+\frac{7}{4}\xi B_e^2  \quad
n_{\bot}=1+\xi B_e^2 \\
v^2_{\parallel}=(1-\frac{7}{2}\xi B_e^2),\quad
v^2_{\bot}=(1-2\xi B_e^2),
\end{align}
for strong magnetic field we have the electric permittivity  and magnetic permeability are

\begin{align}\label{epsilon3}
\epsilon_{11}= (\mu^{-1})_{11}\simeq 1-\frac{\alpha}{3\pi}\left [ {\rm ln}\left (\frac{B_e}{B_c}\right )\right], \\
\epsilon_{33}\simeq 1-\frac{\alpha}{3\pi}\left [ {\rm ln} \left (\frac{B_e}{B_c}\right ) -\frac{B_e}{B_c} \right ],\\
\mu_{33}\simeq 1-\frac{\alpha}{3\pi}\left [{\rm ln} \left (\frac{B_e}{B_c}\right )+1 \right ],
\end{align}

and

\begin{align}
n_{\parallel}\simeq \left(1+\frac{\alpha}{6\pi}\frac{B_e}{B_c} \right),
\quad
n_{\bot}\simeq 1+\frac{\alpha}{6\pi},\\
v^2_{\parallel}\simeq \left(1-\frac{\alpha}{3\pi}\frac{B_e}{B_c} \right),\quad v^2_{\bot}\simeq \left(1-\frac{\alpha}{3\pi} \right).
\end{align}

In the  weak  magnetic field limit for photon propagation perpendicular to the external magnetic field \cite{Elizabeth,Hugo1} we obtain the dispersion equation
\begin{align}
\omega^{WF,(2)}&\simeq\mid\boldsymbol{k_\perp}\mid \left(1-\frac{7}{4}\xi \boldsymbol{B}_e^2 \right).\nonumber\\
\omega^{WF,(3)}&\simeq\mid\boldsymbol{k_\perp}\mid (1-\xi \boldsymbol{B}_e^2)\label{Bperk23}
\end{align}

\noindent while for strong magnetic field limit the result is

\begin{align}\label{omegaS}
\omega^{SF,(2)}
%=\mid\boldsymbol{k}\mid
%(1-\frac{\alpha}{3\pi}
%(\boldsymbol{\hat{b}\times\hat{k}}))
&\simeq \mid\boldsymbol{k_\perp}\mid \left (1-\frac{\alpha}{6\pi}\frac{B_e}{B_c}\right ),\nonumber\\
\omega^{SF,\,(3)}&\simeq\mid \boldsymbol{k_\perp}\mid
%(1-\frac{\alpha}{6\pi}\frac{B}{B_c} (\boldsymbol{\hat{b}\times\hat{k}}))
%=\mid\boldsymbol{k}\mid
\left (1-\frac{\alpha}{6\pi}\right).
\end{align}

\end{document}